\documentclass[]{iopart}
\usepackage{iopams, graphicx, hyperref}

\newcommand{\C}{\mathcal{C}}
\newcommand{\F}{\mathcal{F}}
\newcommand{\N}{\mathcal{N}}
\newcommand{\half}{\case{1}{2}}
\newcommand{\spartial}{{\hat \partial}}
\newcommand{\st}{{\hat t}}
\renewcommand{\Re}{\mathrm{Re}}

\begin{document}

\title[Stable radiation-controlling boundary conditions]
      {Stable radiation-controlling boundary conditions 
       for the generalized harmonic Einstein equations}
\author{O Rinne}
\address{Theoretical Astrophysics 130-33, California Institute of
   Technology, Pasadena, CA 91125, USA}
\ead{rinne@tapir.caltech.edu}

\begin{abstract}
This paper is concerned with the initial-boundary value problem for the 
Einstein equations in a first-order generalized harmonic formulation. 
We impose boundary conditions that preserve 
the constraints and control the incoming gravitational radiation by 
prescribing data for the incoming fields of the Weyl tensor. 
High-frequency perturbations about any given spacetime (including a
shift vector with subluminal normal component) are analyzed using the
Fourier-Laplace technique. We show that the system is boundary-stable.
In addition, we develop a criterion that can be used to detect weak
instabilities with polynomial time dependence, and we show that our system
does not suffer from such instabilities.
A numerical robust stability test supports our claim that the 
initial-boundary value problem is most likely to be well-posed even if
nonzero initial and source data are included.
\end{abstract}


\section{Introduction}

Most attempts to solve the Einstein equations numerically are based on the
Cauchy, or $3+1$ formulation of general relativity, in which one foliates 
spacetime into three-dimensional spacelike hypersurfaces of constant time
$t$. Initial data satisfying the constraint equations are prescribed on the 
$t = 0$ surface and the evolution equations are integrated towards the future.
Typically, one truncates the domain of integration and only aims to
find solutions on a compact spatial manifold with artificial timelike 
boundaries, on which boundary conditions must be specified.
These should satisfy a number of requirements: 

\begin{enumerate}
  \item The initial-boundary value problem (IBVP) should be well-posed, 
    i.e., for given initial and boundary data a unique solution 
    should exist and it should depend continuously on the data. 
  \item The boundary conditions must be compatible with the constraints in 
    the sense that if the constraints are satisfied initially then they 
    remain satisfied at all times.
  \item The boundary conditions should (in some sense) control the incoming
    gravitational radiation. 
\end{enumerate}
The last requirement is of particular importance if one aims to obtain 
reliable information 
about the gravitational radiation emitted by compact sources such as 
coalescing binary black holes. One way to describe the incoming radiation is
in terms of the incoming fields of the evolution system that the Weyl tensor
obeys by virtue of the Bianchi identies. These incoming fields are
proportional to the Newman-Penrose scalar $\Psi_0$, a quantity that is
invariant under infinitesimal coordinate transformations of Kerr
spacetime. If the outer boundary is placed sufficiently far away from the
source then one expects 
\begin{equation}
  \label{eq:zeropsi0}
  \Psi_0 = 0
\end{equation}
 to be a reasonable ``no-incoming-radiation'' boundary condition. 
Conditions of this type have been considered by a number of authors 
\cite{FriedrichNagy,BardeenBuchman,KidderBC,SarbachTiglio}.

Since the initial-boundary problem for Einstein's equations was first
studied by Stewart \cite{Stewart} with numerical relativity in mind,
this has been a very active field of research.
However, there is currently only one formulation, due to Friedrich and 
Nagy \cite{FriedrichNagy}, that satisfies all the above
requirements for the fully nonlinear vacuum Einstein equations. 
It uses a tetrad formalism and evolves the components of the Weyl tensor 
as separate variables. All the boundary conditions are written in maximally
dissipative form and well-posedness of the IBVP follows from standard theorems
for symmetric hyperbolic systems with such boundary conditions \cite{Rauch, 
Secchi1, Secchi2}.
It would be desirable to obtain a similar result for a metric formulation of
the Einstein equations which does not evolve the Weyl tensor separately, such
as most of the formulations currently used in numerical relativity (see
however \cite{BuchmanBardeen} for an implementation of a tetrad-based 
formulation).
In some cases, well-posedness has been proved (at least partially)
for boundary conditions that satisfy conditions (i) and (ii) but not (iii) 
(see for example \cite{IriondoReula,CalabreseLehnerTiglio,
CalabreseCPBC,GundlachMartinGarcia,BonaCPBC,TarfuleaPhD,Alekseenko}, 
in particular harmonic formulations \cite{SzilagyiSchmidtWinicour,
SzilagyiWinicour,BabiucSzilagyiWinicour,KreissWinicour}).
The major obstruction is the fact that the physical boundary 
conditions \eref{eq:zeropsi0} as well as (in general) the 
constraint-preserving boundary conditions contain second rather than first
derivatives of the metric (such boundary conditions are said to be of 
\emph{differential} type). Hence they are not in maximally dissipative
form and the standard theorems do not apply.
Progress can still be made in a certain approximation (typically in
the high-frequency limit or for linearizations about flat space)
using Fourier-Laplace transform methods \cite{GKO,KreissLorenz}. 
This technique yields rather strong necessary conditions for
well-posedness, although sufficient conditions are not known for
boundary conditions that are of differential type.
Applications to the Einstein equations and related problems can be 
found in \cite{Stewart,CalabreseSarbach,ReulaSarbach,SarbachTiglio,RinnePhD,
KreissWinicour}. In particular, boundary conditions satisfying both 
requirements (ii) and (iii) were analyzed in \cite{SarbachTiglio, RinnePhD}.
Necessary conditions for well-posedness were verified but numerical
experiments indicated that the systems still suffered from instabilities.
Using rather different methods based on semigroup theory, Nagy and Sarbach
\cite{NagySarbach} have recently proved well-posedness of the IBVP
(with radiation-controlling boundary conditions) for the linearized 
Einstein equations in the ADM formulation in a certain gauge.
The gauge condition arises from a minimization principle and involves a 
fourth-order elliptic equation for the 
lapse function (which might be expensive to solve numerically). 

In this article, we consider a symmetric hyperbolic formulation of the 
Einstein equations based on generalized harmonic (GH) coordinates 
(see \cite{GH} and references therein).
Such coordinates $x^a$ obey the scalar wave equation
\begin{equation}
  \label{eq:ghgauge}
  \Box x^a = H^a
\end{equation}
with a source $H^a$ that may depend on the coordinates and the spacetime
metric $\psi_{ab}$ (but not on its derivatives). 
In this gauge, the Einstein equations reduce to a system of coupled nonlinear
wave equations, with principal parts
\begin{equation}
  \label{eq:gh2eveq}
  \Box \psi_{ab} \simeq 0.
\end{equation}
Recently, harmonic coordinates have played an important role in the
first successful simulations of the inspiral, merger and ringdown of
binary black holes by Pretorius \cite{Pretorius1, Pretorius2}. 
In \cite{GH}, a first-order formulation of this system was derived,
with particular emphasis on controlling the constraints in numerical 
evolutions.
Boundary conditions satisfying requirements (ii) and (iii) above were
constructed, and a partial result on the well-posedness of these boundary
conditions was stated. The purpose of this paper is to elaborate on and
generalize this result and to provide both analytical and numerical evidence
that the system at hand is a strong candidate for a well-posed
initial-boundary value formulation of the Einstein equations.

In section \ref{sec:IBVP}, we review the evolution equations 
and boundary conditions derived in \cite{GH}.
The Fourier-Laplace analysis of the initial-boundary value problem is 
carried out in section \ref{sec:fl}.
Whereas the well-posedness result in \cite{GH} assumed that the shift 
vector was tangential at the boundary, we lift that restriction here
and allow for an arbitrary shift. 
Because several characteristic fields propagate along the normal
lines of the spacetime foliation, this may cause additional characteristic
fields to be incoming, which complicates the analysis. Nevertheless, 
the result stated in \cite{GH} carries over to the general case.
(We still assume for the analysis that the normal component of the shift 
is subluminal at the boundary, i.e., we disregard situations where
either all the modes are outgoing or all the modes are ingoing.
If all the modes are outgoing, e.g.~inside a black hole, no boundary 
conditions are to be imposed. See section \ref{sec:num} and \cite{GH} 
for numerical tests of the GH evolution system involving this situation.)

The usual necessary condition for well-posedness in the Fourier-Laplace
framework amounts to verifying that a certain complex determinant has no zeros
with positive real part. We strengthen this result by showing that our system
also obeys the \emph{Kreiss condition} \cite{GKO}, which is stronger
and implies that the solution can be estimated in terms of the boundary data.
Related concepts of well-posedness are briefly discussed.
It has been claimed \cite{ReulaSarbach,SarbachTiglio} that even if the Kreiss
condition is satisfied, \emph{weak instabilities} with milder than 
exponential time dependence might exist if nontrivial initial data are
included.
In section \ref{sec:weakinstab}, we show rather generally that the Kreiss
condition excludes any instabilities with polynomial time dependence.
We also demonstrate that for a bad choice of gauge boundary conditions 
that do not satisfy the Kreiss condition, a weak instability does exist.
In order to supplement our analytical results, we perform a series of numerical
\emph{robust stability tests} in section \ref{sec:num}. This involves
adding random data to the initial and boundary conditions and to the
right-hand-sides of the evolution equations. The background
spacetime is taken to be either flat space (including a shift) 
or Schwarzschild.
A summary of the results and conclusions are given in section \ref{sec:concl}.


\section{The initial-boundary value problem}
\label{sec:IBVP}

We begin by presenting the initial-boundary value problem for the
Einstein equations in generalized harmonic gauge that is considered in
this paper. This section follows closely \cite{GH}. Some additional
details concerning the constraint-preserving boundary conditions  
in the case of a non-tangential shift at the boundary are provided.

\subsection{Evolution equations}

We consider the first-order formulation of the generalized harmonic
evolution equations \eref{eq:gh2eveq} derived in \cite{GH}. 
The fundamental variables
are the spacetime metric $\psi_{ab}$ and its first derivatives
$\Phi_{iab} \equiv \partial_i \psi_{ab}$ and 
$\Pi_{ab} \equiv -t^c \partial_c \psi_{ab}$.
Here $t_a \propto \partial_a t$ denotes the unit timelike normal to 
the $t =\mathrm{const.}$ hypersurfaces. 
Lower-case Latin indices from the beginning of the alphabet denote 
4-dimensional spacetime quantities, while 
lower-case Latin indices from the middle of the alphabet are spatial.

The evolution equations take the form
\begin{eqnarray}
  \label{eq:dtpsi}
  \partial_t \psi_{ab} \simeq (1 + \gamma_1) N^k \partial_k \psi_{ab}, \\
  \label{eq:dtpi}
  \partial_t \Pi_{ab} \simeq N^k \partial_k \Pi_{ab} - N g^{ki}
      \partial_k \Phi_{iab} + \gamma_1 \gamma_2 N^k \partial_k
      \psi_{ab}, \\
  \label{eq:dtphi}    
  \partial_t \Phi_{iab} \simeq N^k \partial_k \Phi_{iab} - N
      \partial_i \Pi_{ab} + N \gamma_2 \partial_i \psi_{ab},
\end{eqnarray}
where $\simeq$ indicates that only the principal parts of the
equations are displayed. 
The spatial metric $g_{ij}$, lapse function $N$ and shift vector $N^i$
are related to the 4-metric via the standard ADM form of the line element,
\begin{equation}
  \rmd s^2 = \psi_{ab} \rmd x^a \rmd x^b 
  = - N^2 \rmd t^2 + g_{ij}(\rmd x^i + N^i \rmd t)(\rmd x^j + N^j \rmd t).
\end{equation}

From now on we will choose $\gamma_1 = -1$, which ensures
that the evolution equations are linearly degenerate.
The parameter $\gamma_2$ was introduced in \cite{GH} in order to damp 
violations of the constraint 
\begin{equation}
  \label{eq:C3}
  \C_{iab} \equiv \partial_i \psi_{ab} - \Phi_{iab} = 0. 
\end{equation}
There is also a parameter $\gamma_0$ hidden in the source terms, which
is designed to damp violations of the generalized harmonic gauge 
constraint \eref{eq:ghgauge}, based on a suggestion in \cite{Gundlach}. 
It will play no role in this discussion.

The system (\ref{eq:dtpsi}--\ref{eq:dtphi}) is symmetric hyperbolic
for any choice of parameters $\gamma_1$ and $\gamma_2$.
Hence the Cauchy problem is well-posed \cite{KreissLorenz}.
The characteristic variables and speeds in the direction of a unit spacelike
vector $n_i$ are given by
\begin{eqnarray}
  u^0_{ab} = \psi_{ab}, & \mathrm{speed}\,\, 0, \\
  \label{eq:u1m}
  u^{1\pm}_{ab} = \Pi_{ab} \pm \Phi_{nab} - \gamma_2 \psi_{ab}, \quad &
     \mathrm{speed} -N^n \pm N, \\
  u^2_{Aab} = \Phi_{Aab}, & \mathrm{speed} -N^n,
\end{eqnarray}
where here and in the following, an index $n$ denotes contraction with
$n_i$ (e.g., $v_n = n_i v^i$) and upper-case Latin indices denote 
projection with $P_{ij} \equiv g_{ij} - n_i n_j$ (e.g., $v_A = P_{Ai} v^i$).

We consider a finite spatial domain of integration $\Omega$ with
smooth boundary $\partial \Omega$.
The spatial coordinate location of the boundary remains fixed in time,
i.e., $\partial / \partial t$ is tangential to the
3-dimensional timelike boundary $\partial \Omega \times [0, \infty)$.
Boundary conditions have to be prescribed for the incoming characteristic 
fields, where now $n_i$ refers to the outward-pointing unit spacelike 
normal to $\partial \Omega$.
Note that the number of incoming fields depends on the value of the
normal component $N^n$ of the shift at the boundary. 
For $-N < N^n \leqslant 0$, only the fields $u^{1-}_{ab}$ are incoming;
for $0 < N^n \leqslant N$, the fields $u^{1-}_{ab}$ and $u^2_{Aab}$
are incoming.

\subsection{Constraint-preserving boundary conditions}
\label{sec:CPBC}

The boundary conditions should ensure that if the constraints vanish
initially then they vanish at all times. In other words, the
subsidiary evolution system that the constraints obey as a consequence
of the main evolution equations has to be well-posed, with the unique
solution being the trivial one.

In our case, the primary constraints are the harmonic gauge constraint
\eref{eq:ghgauge}, which can be written in the form
\begin{equation}
  \label{eq:C1}
  \C_a \equiv \psi^{bc} \Gamma_{abc} + H_a = 0
\end{equation}
($\Gamma_{abc}$ being the Christoffel symbols of the metric $\psi_{ab}$),
and the definition constraint $\C_{iab}$ \eref{eq:C3}. 
If we define the first-order constraint variables
\begin{eqnarray}
  \C_{ijab} = 2 \partial_{[i} \Phi_{j]ab} = 2 \partial_{[i} \C_{j]ab},\\
  \label{eq:Fdef}
  \F_a = -t^c \partial_c \C_a + \ldots, \\
  \label{eq:C2def}
  \C_{ia} = \partial_i \C_a - g^{jk} \C_{ijka} + \half g_a{}^j
      \psi^{cd} \C_{ijcd} + \ldots
\end{eqnarray}
(omitting terms proportional to $\C_{iab}$),
then the constraint evolution system can be written in the simple 
form \cite{GH} 
\begin{eqnarray}
  \label{eq:c1ev}
  \partial_t \C_a \simeq 0,\\
  \partial_t \F_a \simeq N^i \partial_i \F_a + N g^{ij} \partial_i \C_{ja},\\
  \partial_t \C_{ia} \simeq N^j \partial_j \C_{ia} + N \partial_i \F_a,\\
  \label{eq:c3ev}
  \partial_t \C_{iab} \simeq 0,\\
  \label{eq:c4ev}
  \partial_t \C_{ijab} \simeq N^k \partial_k \C_{ijab}.
\end{eqnarray}

This system is clearly symmetric hyperbolic, and the characteristic
variables and speeds are
\begin{eqnarray}
  c^{0\pm}_a = \F_a \mp \C_{na}, \quad & \mathrm{speed} -N^n \pm N,\\
  c^1_a = \C_a, & \mathrm{speed}\,\, 0,\\ 
  c^2_{Aa} = \C_{Aa}, & \mathrm{speed} -N^n,\\
  c^3_{iab} = \C_{iab}, & \mathrm{speed}\,\, 0,\\
  c^4_{ijab} = \C_{ijab}, & \mathrm{speed} -N^n.
\end{eqnarray}

Consider first the case $-N < N^n \leqslant 0$.
The only incoming constraint fields are $c^{0-}_a$. 
We impose completely absorbing boundary conditions on them:
\begin{equation}
  \label{eq:c0mbc}
  c^{0-}_a \doteq 0
\end{equation}
($\doteq$ denoting equality at the boundary).
These can be translated into conditions on the normal derivatives of 
four of the main incoming fields $u^{1-}_{ab}$ by noting that
\begin{eqnarray} \fl
  c^{0-}_a \simeq  \sqrt{2} \left[ k^{(c} \psi^{d)}{}_a - \half k_a
    \psi^{cd} \right] \partial_n u^{1-}_{cd} \nonumber\\
    + \half P^A{}_a \psi^{cd} \partial_A \Pi_{cd} - P^{AB} \partial_A \Pi_{Ba}
    - P^{AB} t^b \partial_A \Phi_{Bba} \\
    + \half t_a \psi^{cd} P^{AB} \partial_A \Phi_{Bcd}
    + \gamma_2 (P^{AB} \partial_A \psi_{Ba} - \half P^A{}_a \psi^{cd} 
    \partial_A \psi_{cd} ) \nonumber\\
    + P^{AB} \partial_A \Phi_{nBa} - \half P^A{}_a \psi^{cd}
    \partial_A \Phi_{ncd},\nonumber
\end{eqnarray}
where we have introduced the ingoing null vector 
$k^a = (t^a - n^a)/\sqrt{2}$.

In other words, the constraint-preserving boundary conditions imply
boundary conditions on the normal derivatives of the following
projection of $u^{1-}_{ab}$,
\begin{equation}
  P^{(\mathrm{C})cd}_{ab} u^{1-}_{cd} \equiv \left[ \half P_{ab} P^{cd} - 2
    l_{(a} P_{b)}{}^{(c} k^{d)} + l_a l_b k^c k^d \right] u^{1-}_{cd},
\end{equation}
where we have also introduced the outgoing null vector 
$l^a = (t^a + n^a)/\sqrt{2}$.
This projection has four degrees of freedom. Of the remaining six
degrees of freedom of $u^{1-}_{ab}$, two will be fixed by the
physical boundary conditions and four by the gauge boundary conditions.

Next, we consider the case $0 < N^n \leqslant N$.
Now the fields $c^2_{Aa}$ and $c^4_{ijab}$ are incoming as well.
In addition to \eref{eq:c0mbc}, we impose the boundary conditions 
\begin{equation}
  \label{eq:c4bc}
  0 \doteq c^4_{nAbc} \simeq \partial_n u^2_{Abc} - \partial_A \Phi_{nbc} ,
\end{equation}
which are conditions on the normal derivatives of the main characteristic
fields $u^2_{Aab}$.
The remaining incoming constraint fields $c^2_{Aa}$ and $c^4_{ABab}$
also need boundary conditions. However, together with the physical and
gauge boundary conditions,
we have already used up all the incoming modes of the main evolution 
system and cannot impose any further boundary conditions actively. 
Fortunately, it turns out that $c^2_{Aa} \doteq 0$ and $c^4_{ABab} \doteq 0$ 
as a consequence of the constraint-preserving boundary conditions we 
have already imposed and the evolution equations. We prove this 
important point in the appendix, using the Fourier-Laplace method 
(we only consider the limit of high-frequency perturbations about a 
fixed background spacetime in that part of the analysis).

To summarize, we have imposed homogeneous maximally dissipative
boundary conditions for the constraint evolution system. Because of general 
theorems for quasilinear symmetric hyperbolic systems with such boundary
conditions \cite{Rauch, Secchi1, Secchi2}, it follows that the constraint
evolution system is well-posed and the unique solution is the trivial one.

Setting to zero the incoming fields of the constraint evolution system
at the boundary as in \eref{eq:c0mbc} is the standard prescription used
in most works on constraint-preserving boundary conditions 
(e.g., \cite{Stewart,IriondoReula,CalabreseLehnerTiglio,KidderBC,BonaCPBC}).
This ensures that any constraint violations generated in a numerical 
evolution leave the computational domain without reflections
(at least for normal incidence). 
In harmonic formulations, one sometimes considers the simpler alternative
$\C_a \doteq 0$ \cite{KreissWinicour}, which does not involve any
derivatives of the fundamental fields.
These Dirichlet boundary conditions are clearly also consistent 
with the constraints but they constitute a reflective boundary condition for
the constraint evolution system (\ref{eq:c1ev}--\ref{eq:c4ev}). 
Numerical tests of such boundary conditions for an axisymmetric 
version \cite{RinneStewart} of the Z4 system \cite{BonaZ4} 
(which is very similar \cite{Gundlach} to the generalized harmonic 
evolution system considered here) 
indicate that they can cause significant reflections of constraint 
violations \cite{RinnePhD}.

\subsection{Physical boundary conditions}

The physical boundary conditions used in \cite{GH} are designed to control 
the gravitational radiation entering the domain through the boundary,
following the prescription used 
in \cite{FriedrichNagy,BardeenBuchman,KidderBC,SarbachTiglio}.
As explained in the introduction, the incoming radiation can be 
described in terms of the ingoing characteristic fields $w^-_{ab}$ 
of the Weyl tensor evolution system,
\begin{equation}
  w^-_{ab} = 2 (P_a{}^c P_b{}^d - \half P_{ab} P^{cd}) k^e k^f C_{cedf}.
\end{equation}
(The fields $w^-_{ab}$ are proportional to the Newman-Penrose scalar
$\Psi_0$ for a null tetrad containing the null vectors $k^a$
and $l^a$ defined above.)
As a boundary condition, we impose
\begin{equation}
  \label{eq:physbc}
  w^-_{ab} \doteq \partial_t h^{(\mathrm{P})}_{ab} .
\end{equation} 
The data $h^{(\mathrm{P})}_{ab}$ may be used to inject a gravitational 
wave through the boundary, for instance.
Note that the expression for $w^-_{ab}$ is only unique up to multiples
of the constraints $\C_{ijab}$ related to the index ordering in the 
first-order reduction. For a particular choice of index ordering, 
we can write $w^-_{ab}$ in the form
\begin{eqnarray}
  \label{eq:weylexpr} \fl
  w^-_{ab} \simeq (P_a{}^A P_b{}^B - \half P_{ab} P^{AB}) & \left[ 
    \partial_n (u^{1-}_{AB} + \gamma_2 u^{0}_{AB}) 
    \right. \nonumber\\& 
    + 2 g^{ij} \partial_A \Phi_{ijB} - g^{ij} \partial_A \Phi_{Bij} -
    P^{CD} \partial_C \Phi_{DAB} 
    \\&  \left.
    + t^c \partial_A \Phi_{nBc} - \partial_A \Pi_{Bn} 
    - t^c \partial_A \Phi_{Bnc} \right] .\nonumber
\end{eqnarray}
Hence \eref{eq:physbc} provides a boundary condition on the normal
derivatives of
\begin{equation}
  P^{(\mathrm{P})cd}_{ab} u^{1-}_{cd} \equiv
     (P_a{}^c P_b{}^d - \half P_{ab} P^{cd}) u^{1-}_{cd}.
\end{equation}

In \cite{BuchmanSarbach}, a hierarchy of boundary conditions has
recently been developed that are perfectly absorbing for linearized
gravitational radiation up to some arbitrary angular momentum number $L$.
They can be viewed as successive improvements of the $\Psi_0 \doteq 0$
boundary conditions and take the form
\begin{equation}
  \label{eq:improvedphysbc}
  \partial_t (r^2 l^a \partial_a)^{L-1} (r^5 \Psi_0) \vert_{r = R} = 0,
\end{equation}
where $r$ is an areal radial coordinate and it is assumed that the
boundary is a sphere of radius $R$.
We remark here that the stability result derived in this paper carries
over to these improved boundary conditions (cf.~section \ref{sec:ghfl}).

\subsection{Gauge boundary conditions}

The remaining components of $u^{1-}_{ab}$ correspond to gauge
degrees of freedom. To understand this, one should note that the
generalized harmonic condition \eref{eq:ghgauge} does not fix the 
coordinates completely: there is still a remaining gauge freedom
$x^a \rightarrow x^a + \xi^a$ for infinitesimal coordinate displacements 
$\xi^a$ that satisfy the wave equation. We may exploit this gauge freedom 
in order to choose the four remaining boundary conditions in any way we like.
Our viewpoint here is that the gauge boundary conditions should guarantee
that the IBVP is well-posed, while still admitting 
arbitrary boundary data in order to be able to impose a variety of gauge
conditions.

The simplest way appears to prescribe data $h_{ab}^{(\mathrm{G})}$ 
for those components of $u^{1-}_{ab}$ that are not specified by the
constraint-preserving and physical boundary conditions,
\begin{eqnarray}
  \label{eq:gaugebc}
  P^{(\mathrm{G})cd}_{ab} u^{1-}_{cd} &\equiv \left[ \delta_{(a}^c 
    \delta_{b)}^d 
  - P^{(\mathrm{C})cd}_{ab} - P^{(\mathrm{P})cd}_{ab} \right] u^{1-}_{cd}
     \nonumber\\&
   = \left[ 2 l_{(a}k_{b)} l^c k^d + k_a k_b l^c l^d - 2 k_{(a}
     P_{b)}{}^c l^d \right] u^{1-}_{cd} \doteq h^{(\mathrm{G})}_{ab}.
\end{eqnarray}
Equivalently, we can write \eref{eq:gaugebc} in the simpler form
\begin{equation}
  \label{eq:gaugebc1}
  l^b u^{1-}_{ab} \doteq h^{(\mathrm{G})}_a.
\end{equation}
These (with $h^{(\mathrm{G})}$ = 0) are the gauge boundary conditions
used in the numerical implementation of \cite{GH}.

For illustrational purposes only, we shall also consider the following
alternative set of gauge boundary conditions, which will turn out to
be ill-posed (to varying extent):
\begin{equation}
  \label{eq:altgaugebc}
  t^b \Pi_{ab} \doteq h^{(\mathrm{G'})}_{a} .
\end{equation}
To linear order, these amount to prescribing (the time derivative of)
the lapse and shift at the boundary.


\section{Fourier-Laplace analysis}
\label{sec:fl}

In this section, we discuss necessary conditions for well-posedness
using the Fourier-Laplace technique.
We begin with a review of the general theory, mainly following 
\cite{Stewart, GKO} and highlighting some open questions.
We then apply it to the generalized harmonic evolution system for the case 
of high-frequency perturbations about any arbitrary spacetime.

\subsection{General theory}
\label{sec:genfl}

Consider a linear symmetric hyperbolic system of evolution equations
\begin{equation}
  \label{eq:geneveqs}
  \partial_t \bi u = A^i \partial_i \bi u,
\end{equation}
where $\bi u$ is an $m$-dimensional vector, the $A^i$ are $m\times
m$ constant symmetric matrices, and $1 \leqslant i \leqslant n$.
We take the spatial domain of integration to be the quarter-space 
$t \geqslant 0, x^1 \geqslant 0$, $-\infty < x^2, x^3, \ldots , x^n < \infty$.
Initial data are prescribed at $t = 0$,
\begin{equation}
  \label{eq:geninidat}
  \bi u(0, x^i) = \bi f(x^i).
\end{equation}
Let $l$ denote the number of negative eigenvalues of $A^1$ (i.e., the 
number of incoming modes).
We impose boundary conditions at $x^1 = 0$ of the form
\begin{equation}
  \label{eq:genbcs}
  S^i \partial_i \bi u \doteq \partial_t \bi h,
\end{equation}
where the $S^i$ are constant $l \times m$ matrices 
and $\bi h$ is a vector of boundary data (this non-standard
form of the boundary conditions is required for our application in 
section \ref{sec:ghfl}).

We allow for the boundary to be (uniformly) characteristic, which is 
the case for most applications in physics, including the system discussed 
in this paper. After a suitable orthogonal transformation of the 
variables, we may write
\begin{equation}
  \label{eq:Apartition}
  A^1 = \left( \begin{array}{cc} \mathbb{O}_k & 0 \\ 0 & A \end{array} \right),
\end{equation}
where $\mathbb{O}_k$ is the $k\times k$ zero matrix and $A$ is an 
$(m-k)\times (m-k)$ matrix. Similarly, we split
\begin{equation}
  \bi u = (\bi z, \bi v)^T .
\end{equation}

The initial-boundary value problem (\ref{eq:geneveqs}--\ref{eq:genbcs})
can be solved by means of a Laplace transform with respect to time
and a Fourier transformation with respect to the spatial coordinates
$x^A$ tangent to the boundary ($2 \leqslant A \leqslant n$), 
i.e., we write $\bi u$ as a superposition of modes
\begin{equation}
  \label{eq:genflansatz}
  \bi u(t, x^i) = \tilde \bi u (x^1) \exp(s t + \rmi \omega_A x^A)
\end{equation}
with $s \in \mathbb{C}$, $\Re(s) > 0$ and $\omega_A \in \mathbb{R}$. 
By inserting \eref{eq:genflansatz} into the evolution equations 
\eref{eq:geneveqs}, we obtain the Fourier-Laplace-transformed version
\begin{equation}
  \label{eq:genflt}
  s \tilde \bi u = A^1 \partial_1 \tilde \bi u 
  + \rmi \omega_A A^A \tilde \bi u.
\end{equation}
Partitioning the matrix 
\begin{equation}
  \label{eq:flmatrixB}
  s \mathbb{I}_m - \rmi \omega_A A^A \equiv B
  = \left( \begin{array}{cc} B_{11} & B_{12} \\ B_{12}{}^T & B_{22} 
    \end{array} \right)
\end{equation}
in the same fashion as in \eref{eq:Apartition} ($\mathbb{I}_m$ being
the $m\times m$ unit matrix), equation \eref{eq:genflt} splits into
\begin{eqnarray}
  \label{eq:genfltalg}
  0 &=& B_{11} \tilde \bi z + B_{12} \tilde \bi v,\\
  A \partial_1 \tilde \bi v &=& B_{12}{}^T \tilde \bi z + B_{22}
    \tilde \bi v.
\end{eqnarray}
Because the $A^i$ are symmetric, the matrix $B$ and in particular
$B_{11}$ in \eref{eq:flmatrixB} has only nonzero eigenvalues. 
Hence we may use the algebraic relations \eref{eq:genfltalg} in 
order to eliminate $\tilde \bi z$ in favour of $\tilde \bi v$. 
We are left with the system of ordinary differential equations (ODEs)
\begin{equation}
  \label{eq:genodes}
  \partial_1 \tilde \bi v = M \tilde \bi v, \qquad
  M \equiv A^{-1} (B_{22} - B_{12}{}^T B_{11}{}^{-1} B_{12}).
\end{equation}

A simple argument \cite{MO, GKO} shows that $M$ has precisely $l$ 
eigenvalues $\lambda_1, \lambda_2, \ldots, \lambda_l$ 
with negative real part. Let $\bi w_1, \bi w_2, \ldots, \bi w_l$
denote the corresponding eigenvectors.
Then the general $L^2$ solution of \eref{eq:genodes} is given by
\begin{equation}
  \label{eq:gendecsoln}
  \tilde \bi v(s, x^1, \omega) = \sum_{j=1}^l \sigma_j \bi w_j (s,
  \omega) \exp(\lambda_j x^1)
\end{equation}
with complex integration constants $\sigma_j$.
(If the eigenvectors do not span the space then some of the $\sigma_j$ 
have to be replaced with polynomials in $x^1$ -- this will occur as a
special case in our application.)

The constants $\sigma_j$ are determined by the boundary conditions:
substituting \eref{eq:gendecsoln} into the Fourier-Laplace transform of 
\eref{eq:genbcs},
\begin{equation}
  \label{eq:genbcflt}
  S^1 \partial_1 \tilde \bi u + \rmi \omega_A S^A \tilde \bi u 
  \doteq s \tilde \bi h,
\end{equation}
we obtain a system of linear equations
\begin{equation}
  \label{eq:genCsys}
  C(s, \omega) \bsigma = \tilde \bi h,
\end{equation}
where $C$ is an $l\times l$ matrix.

Consider first the homogeneous boundary conditions, $\tilde \bi h = 0$.
Suppose that $\det C = 0$ for some $s$ with $\Re(s) > 0$.
Then \eref{eq:genCsys} has a nontrivial solution, and hence the
IBVP (\ref{eq:geneveqs}--\ref{eq:genbcs}) with homogeneous boundary
conditions has a non-trivial solution of the form \eref{eq:genflansatz}.
Now observe that with \eref{eq:genflansatz},
\begin{equation}
  \bi u(t, x^i) = \tilde \bi u(\alpha x^1) \exp(\alpha s t + \rmi \alpha
  \omega_A x^A)
\end{equation}
is also a solution, for any $\alpha > 0$. Thus we can find solutions
which grow exponentially at an arbitrarily fast rate, and the IBVP is
ill-posed. Such solutions are called \emph{strong instabilities}.
(In a numerical simulation, $\alpha$ is determined by the
highest frequency that can be represented on the grid. Hence the
growth rate of the instability increases with resolution.)
We conclude that the \emph{determinant condition}
\begin{equation}
  \label{eq:gendetcond}
  \det C(s, \omega) \neq 0 \quad \mathrm{for} \quad \Re(s) > 0
\end{equation}
is a necessary condition for well-posedness.

Next, we consider the inhomogeneous boundary conditions \eref{eq:genbcs}.
Formally, we can solve \eref{eq:genCsys} for the integration
constants $\bsigma$ provided that the determinant condition is satisfied.
What remains to be shown is that the solution \eref{eq:gendecsoln} 
can be bounded in terms of the boundary data, 
\begin{equation}
  \label{eq:genkreisscond1}
  | \tilde \bi v(s,0,\omega) | \leqslant K | \tilde \bi h(s, \omega) | ,
\end{equation}
with a constant $K > 0$ that is independent of $s$ and $\omega$.
This is known as the \emph{Kreiss condition} \cite{GKO}. 
Provided that the eigenvectors $\bi w_i$ in \eref{eq:gendecsoln} are 
normalized in such a way that they remain finite as $\Re(s) \downarrow 0$ 
and as $|s| \rightarrow \infty$, \eref{eq:genkreisscond1} is equivalent to
\begin{equation}
  \label{eq:genkreisscond2}
  \det C(s, \omega) \neq 0 \quad \mathrm{for} \quad \Re(s) \geqslant 0.
\end{equation}
Comparing this with \eref{eq:gendetcond}, the additional requirement
is that there be no zeros $s$ of $\det C$ with $\Re(s) = 0$.
Such zeros are known as \emph{generalized eigenvalues}.

If the Kreiss condition is satisfied then it follows
immediately that the IBVP (\ref{eq:geneveqs}, \ref{eq:genbcs}) 
is well-posed for vanishing initial data. 
The unique solution is given by the inverse of the Fourier-Laplace transform, 
which is well-defined because of the bound \eref{eq:genkreisscond1}.
In particular, we obtain an estimate of the form
\begin{equation} 
  \label{eq:bdrystabestimate}
  \int_0^t \| \bi u(\cdot,\tau) \|_\Omega^2 \, \rmd\tau 
  \leqslant K_T \int_0^t \| \bi h(\cdot,\tau) \|_{\partial \Omega}^2 \, \rmd\tau
\end{equation}
in every time interval $0 \leqslant t \leqslant T$, where the constant
$K_T$ is independent of the data $\bi h$ (the norms are $L^2$ norms
over the half-space and the boundary, respectively).
We say that the system is \emph{boundary-stable} \cite{KreissWinicour}. 
It is this definition of stability that we shall prove for the generalized
harmonic evolution system (section \ref{sec:ghfl}).

Closely related to boundary stability is the concept of \emph{strong
well-posedness in the generalized sense} \cite{GKO, KreissLorenz, 
KreissWinicour}.
This requires that if we add a source term $\bi F(t, x^i)$ to the
right-hand-side of the evolution equations \eref{eq:geneveqs}, the
following estimate holds,
\begin{eqnarray}
  \label{eq:genwpestimate} \fl
  \int_0^t \| \bi u(\cdot,\tau) \|_\Omega^2 \, \rmd\tau 
  + \int_0^t \| \bi v(\cdot,\tau) \|_{\partial \Omega}^2 \, \rmd\tau 
  \nonumber\\
  \leqslant K_T \left( \int_0^t \| \bi h(\cdot,\tau) \|_{\partial
      \Omega}^2 \, \rmd\tau
  + \int_0^t \| \bi F(\cdot,\tau) \|_\Omega^2 \, \rmd\tau \right)
\end{eqnarray}
(recall that $\bi v$ refers to the modes with non-zero speeds).
The crucial ingredient in proving \eref{eq:genwpestimate} given
\eref{eq:bdrystabestimate} is the construction of a 
\emph{symmetrizer} \cite{Kreiss, MO, KreissWinicour}.
In doing this, it is usually assumed that the boundary conditions are
in \emph{maximally dissipative} form
\begin{equation}
  \bi v^- \doteq S \bi v^+ + \bi g,
\end{equation}
where $\bi v^-$ and $\bi v^+$ are the ingoing and outgoing 
(non-zero speed) modes and $S$ is a (for our purposes constant) matrix. 
The symmetrizer method does not appear to be applicable if the
boundary conditions are of the differential type \eref{eq:genbcs}.

So far we have assumed that the initial data $\bi f$ vanish.
One can always treat the case of general initial data by considering e.g.
$\bi u' \equiv \bi u - e^{-t} \bi f$,
so that the problem for $\bi u'$ has zero initial data.
However, in the evolution equations for $\bi u'$ there will be an
additional source term containing \emph{derivatives} of $\bi f$, which will
appear in the estimate generalizing \eref{eq:genwpestimate}.
What one would like instead is an estimate of the form
\begin{eqnarray} \fl
  \label{eq:strongwpestimate}
  \int_0^t \| \bi u(\cdot,\tau) \|_\Omega^2 \, \rmd\tau 
  + \int_0^t \| \bi v(\cdot,\tau) \|_{\partial \Omega}^2 \, \rmd\tau 
  \nonumber\\
  \leqslant K_T \left( \| \bi f(\cdot) \|_\Omega^2 
  + \int_0^t \| \bi F(\cdot,\tau) \|_\Omega^2 \, \rmd\tau
  + \int_0^t \| \bi h(\cdot,\tau) \|_{\partial \Omega}^2 \, \rmd\tau \right),  
\end{eqnarray}
without any derivatives of $\bi f$. 
This is referred to as \emph{strong well-posedness}. 
Majda and Osher \cite{MO} showed in the present case of a uniformly
characteristic boundary that the Kreiss condition is also sufficient
for strong well-posedness provided that the boundary conditions are 
maximally dissipative.
However, nothing is known in general for boundary conditions of the form 
\eref{eq:genbcs}.
In section \ref{sec:weakinstab}, we will argue that the Kreiss condition is
still useful in order to rule out certain \emph{weak instabilities} that have 
a polynomial time dependence.

We close this section with some remarks on the scope of the
Fourier-Laplace method. Technically, one can only apply Fourier and Laplace
transforms if the evolution equations and boundary conditions have 
constant coefficients. However, well-posedness is a concept that is
associated with the behaviour of the high-frequency components of the
solution. In many cases one can show that a problem with variable 
coefficients is strongly well-posed if all problems obtained by freezing the 
coefficients are strongly well-posed (\cite[Theorem 8.4.9]{KreissLorenz}, 
see \cite{Agranovich} for the extension from strictly hyperbolic to symmetric 
hyperbolic systems, see also the discussion in \cite{KreissWinicour}). 
Strong well-posedness can be further extended from linear to quasilinear 
systems (such as the Einstein equations), in which case the estimates will 
in general only hold in a finite time interval (see for 
example \cite[sec.~8.5]{KreissLorenz}, \cite{Secchi1,Secchi2}).
More general spatial domains $\Omega$ than the half-space can easily
be treated by splitting $\partial \Omega$ into a finite
number of portions, each of which can be smoothly mapped to the half-space
\cite[sec.~9.6.2]{GKO}.
One should note, however, that the proofs of the above extensions 
usually assume the boundary conditions to be maximally dissipative -- 
the situation is much less clear for differential boundary conditions.

\subsection{Application to the GH system}
\label{sec:ghfl}

In this section, we apply the Fourier-Laplace technique to the generalized
harmonic system of section \ref{sec:IBVP}. For previous applications of
this method to various formulations of the Einstein equations, 
see e.g.~\cite{Stewart,CalabreseSarbach,ReulaSarbach,SarbachTiglio,RinnePhD,
  KreissWinicour}.

First we construct the most general frozen-coefficient problem
by considering the limit of high-frequency perturbations of an
arbitrary fixed background spacetime.
Consider a point $p$ at the boundary. By rescaling and rotating the
spatial coordinates if necessary, we can achieve that the 3-metric at $p$ is
$g_{ij} = \delta_{ij}$, and by rescaling the time coordinate, we may
also assume that the lapse at $p$ is $N = 1$. However, we have to
allow for an arbitrary shift vector. (If we performed a coordinate
transformation that affected the shift vector, we would have to consider
a moving boundary.)
Furthermore, we may assume that the boundary is located at $x^1 \equiv x = 0$ 
and that the domain of integration is $x > 0$.

Hence the evolution equations (\ref{eq:dtpsi}--\ref{eq:dtphi}) become
\begin{eqnarray}
  \label{eq:dtpsi0}
  \spartial_t \psi_{ab} = N^A \partial_A \psi_{ab}, \\
  \label{eq:dtpi0}
  \spartial_t \Pi_{ab} = N^x \partial_x \Pi_{ab} 
      -\partial^k \Phi_{kab} - \gamma_2 N^k \partial_k \psi_{ab}, \\
  \label{eq:dtphi0}    
  \spartial_t \Phi_{jab} = N^x \partial_x \Phi_{jab} 
      - \partial_j \Pi_{ab} + \gamma_2 \partial_j \psi_{ab},
\end{eqnarray}
where we have introduced the operator $\spartial_t \equiv \partial_t -
N^A \partial_A$, $N^i$ refers to the (constant) background shift vector,
and it is now understood that spatial indices are
raised and lowered with the unit metric.

We consider an ansatz similar to \eref{eq:genflansatz},
\begin{equation}
  \label{eq:flansatz}
  \bi u(t, x^i) = \tilde \bi u(x) \exp \left[ s t 
    + \rmi \omega_A (x^A + N^A t) \right]
\end{equation}
for all the dependent variables 
$\bi u = \{ \psi_{ab}, \Pi_{ab}, \Phi_{iab} \}$, where 
$s \in \mathbb{C}$ with $\Re(s) > 0$ and $x^A = \{ y,z \}$.
Because of the rotational invariance of the evolution equations and
boundary conditions, we may without loss of generality take 
$\omega_A = \omega \delta_A{}^y$, say, with $\omega > 0$
(note that $\omega = 0$ can be excluded because we have taken the
high-frequency limit).
It is convenient to eliminate $\omega$ from the following equations by
defining $\xi \equiv \omega x$ and $\zeta \equiv s/\omega$.

Let us first consider the case $N^x = 0$.
Inserting the ansatz \eref{eq:flansatz} into the evolution equations 
(\ref{eq:dtpsi0}--\ref{eq:dtphi0}), we first obtain the algebraic relations
\begin{eqnarray}
  \zeta \tilde \psi_{ab} = \rmi N^y \tilde \psi_{ab},\\
  \zeta \tilde \Phi_{yab} = -\rmi \tilde \Pi_{ab},\\
  \zeta \tilde \Phi_{zab} = 0.
\end{eqnarray}
From these we deduce that $\tilde \psi_{ab} = \tilde \Phi_{zab} = 0$, and we 
eliminate $\tilde \Phi_{yab}$ in favour of $\tilde \Pi_{ab}$.
The remaining two evolution equations are turned into the ODE system
\begin{equation}
  \partial_\xi \left( \begin{array}{c} \tilde \Pi_{ab} \\ \tilde
      \Phi_{xab} \end{array} \right) 
  = \left( \begin{array}{cc} 0 & -\zeta \\ -\zeta - \frac{1}{\zeta} & 0 
    \end{array} \right)
  \left( \begin{array}{c} \tilde \Pi_{ab} \\ \tilde \Phi_{xab} 
    \end{array} \right) .
\end{equation}
The matrix has eigenvalues 
\begin{equation}
  \lambda_1^\pm = \pm \sqrt{1 + \zeta^2}
\end{equation}
(the branch of the square root is chosen such that $\Re(\lambda_1^+) > 0$
for $\Re(\zeta) > 0$).
Since we are only interested in $L^2$ solutions, i.e., solutions that
decay as $x \rightarrow \infty$, we have to pick the eigenvalue $\lambda_1^-$.
The corresponding eigenvector is $(\zeta, -\lambda_1^-)^T$ and hence the 
general $L^2$ solution is
\begin{equation} 
  \label{eq:zerobetasoln}
  \left( \begin{array}{c} \tilde \Pi_{ab} \\ \tilde \Phi_{xab} \\
    \tilde \Phi_{yab} \\ \tilde \Phi_{zab} \end{array} \right) = 
  \sigma_{1ab}  \, \rme^{\lambda_1^- \xi} \, K_1^{-1}
  \left( \begin{array}{c} \zeta \\ -\lambda_1^- \\ -\rmi \\ 0 \end{array}
  \right) ,
\end{equation}
with arbitrary complex constants $\sigma_{1ab}$. The normalization
constant $K_1$ (and all similar constants in the following) is equal 
to the norm of the vector immediately following it.

Next we discuss the case $\beta \equiv N^x \neq 0$. 
We obtain $\tilde \psi_{ab} = 0$ as before, but now the ODE system reads
\begin{equation} \fl
  \partial_\xi \left( \begin{array}{c} \tilde \Pi_{ab} \\ \tilde \Phi_{xab} \\
      \tilde \Phi_{yab} \\ \tilde \Phi_{zab} \end{array} \right) = 
  \left( \begin{array}{cccc} -\gamma^2 \beta \zeta & -\gamma^2 \zeta &
      -\rmi\gamma^2 \beta & 0 \\ -\gamma^2 \zeta & - \gamma^2 \beta \zeta &
      -\rmi\gamma^2 & 0 \\ \rmi/\beta & 0 & \zeta/\beta & 0 \\
      0 & 0 & 0 & \zeta/\beta \end{array} \right)
  \left( \begin{array}{c} \tilde \Pi_{ab} \\ \tilde \Phi_{xab} \\
      \tilde \Phi_{yab} \\ \tilde \Phi_{zab} \end{array} \right) ,
\end{equation}
where we have introduced the shorthand $\gamma^2 \equiv (1 - \beta^2)^{-1}$.
The eigenvalues are
\begin{equation}
  \lambda_1^\pm = \gamma^2 (-\beta \zeta \pm \sqrt{\zeta^2 + \gamma^{-2}}),
  \qquad \lambda_2 = \frac{\zeta}{\beta}.
\end{equation}

For $\beta > 0$, only $\lambda_1^-$ has negative real part 
(recall that we are assuming $|\beta| < 1$) and the corresponding solution is
\begin{equation}
  \label{eq:posbetasoln}
    \left( \begin{array}{c} \tilde \Pi_{ab} \\ \tilde \Phi_{xab} \\
      \tilde \Phi_{yab} \\ \tilde \Phi_{zab} \end{array} \right) =
    \sigma_{1ab} \, \rme^{\lambda_1^- \xi} \, K_1^{-1}
    \left( \begin{array}{c} \zeta - \beta \lambda_1^- \\ -\lambda_1^- \\ -\rmi
        \\ 0 \end{array} \right) .
\end{equation}
One observes that this contains the above result \eref{eq:zerobetasoln}
for $\beta = 0$ in a regular way so that we do not need to discuss these 
two cases separately in the following.

For $\beta < 0$, the eigenvalue $\lambda_2$ also has 
negative real part and hence the general $L^2$ solution is
\begin{eqnarray} \fl
  \label{eq:negbetasoln} 
    \left( \begin{array}{c} \tilde \Pi_{ab} \\ \tilde \Phi_{xab} \\
      \tilde \Phi_{yab} \\ \tilde \Phi_{zab} \end{array} \right) =&
    \sigma_{1ab} \rme^{\lambda_1^- \xi} \, K_1^{-1}
    \left( \begin{array}{c} \zeta - \beta \lambda_1^- \\ -\lambda_1^- \\ -\rmi
        \\ 0 \end{array} \right) 
    \nonumber\\&
  + \rme^{\lambda_2 \xi} \left[ \sigma_{2ab} \, K_2^{-1}
    \left( \begin{array}{c} 0 \\ -\rmi \beta \\ \zeta \\ 0 \end{array} \right) 
  + \sigma_{3ab} 
    \left( \begin{array}{c} 0 \\ 0 \\ 0 \\ 1 \end{array} \right) \right] .
\end{eqnarray}
More care is needed if $\zeta = -\beta$ because in that case, 
$\lambda_1^- = \lambda_2 = -1$ and the corresponding eigenvectors are 
degenerate (this is the only case in which such a situation occurs).
The most general solution is now of the form
\begin{eqnarray} \fl
  \label{eq:specnegbetasoln} 
    \left( \begin{array}{c} \tilde \Pi_{ab} \\ \tilde \Phi_{xab} \\
      \tilde \Phi_{yab} \\ \tilde \Phi_{zab} \end{array} \right) =
  \rme^{- \xi} &\left[ \sigma_{1ab} \, K_1^{-1}
    \left( \begin{array}{c} -\beta \\ 0 \\ \rmi \\ 0 \end{array} \right)
  + \left( \sigma_{1ab} \xi + \sigma_{2ab} \right) K_2^{-1}
    \left( \begin{array}{c} 0 \\ 1 \\ -\rmi \\ 0 \end{array} \right) 
  + \sigma_{3ab}
    \left( \begin{array}{c} 0 \\ 0 \\ 0 \\ 1 \end{array} \right)  \right] 
  \nonumber\\
\end{eqnarray}

In order to work out the Fourier-Laplace transform of the boundary
conditions, it is useful to note that in our set-up, 
$\{t^a, (\partial/\partial x)^a = -n^a, (\partial/\partial y)^a, 
(\partial/\partial z)^a \}$ 
form an orthonormal basis, and 
$P_{ij} = \delta_i^y \delta_j^y + \delta_i^z \delta_j^z$.
An index $\hat t$ will be used to denote contraction with the timelike
normal $t^a$. We remark that the $\partial_i \psi_{ab}$ terms
hidden in the omitted $\C_{iab}$ terms in (\ref{eq:Fdef}, \ref{eq:C2def})
do not enter in the following because we already know that 
$\tilde \psi_{ab} \equiv 0$.

The Fourier-Laplace transform of the constraint-preserving boundary
conditions \eref{eq:c0mbc} then reads
\begin{eqnarray}
  \label{eq:flc0mbc1} 
\fl  \zeta \tilde h^{(\mathrm{C})}_1 \doteq \tilde c_\st^{0-} = & 
    \partial_\xi \left( - \half \tilde \Pi_{\st\st} - \tilde \Pi_{x\st} 
      - \half \tilde \Pi_{xx} - \half \tilde \Pi_{yy} - \half \tilde \Pi_{zz}
       \right. \nonumber\\ & \qquad  \left.
      - \half \tilde \Phi_{x\st\st} - \tilde \Phi_{xx\st} 
      - \half \tilde \Phi_{xxx} - \half \tilde \Phi_{xyy} 
      - \half \tilde \Phi_{xzz} \right) \\ &
    + \rmi \left( -\tilde \Pi_{y\st} - \half \tilde \Phi_{y\st\st} 
      - \half \tilde \Phi_{yxx} - \half \tilde \Phi_{yyy} 
      - \half \tilde \Phi_{yzz} - \tilde \Phi_{xy\st} \right) , \nonumber
\end{eqnarray}
\begin{eqnarray}
\fl  \zeta \tilde h^{(\mathrm{C})}_2 \doteq \tilde c_x^{0-} =& 
    \partial_\xi \left( -\half \tilde \Pi_{\st\st} - \tilde \Pi_{\st x}
      - \half \tilde \Pi_{xx} + \half \tilde \Pi_{yy} + \half \tilde \Pi_{zz} 
      \right.  \nonumber\\ & \qquad \left.
      - \half \tilde \Phi_{x\st\st} - \tilde \Phi_{x\st x} 
      - \half \tilde \Phi_{xxx} + \half \tilde \Phi_{xyy} 
      + \half \tilde \Phi_{xzz} \right) \\ &
    + \rmi \left( -\tilde \Pi_{yx} - \tilde \Phi_{y\st x} - \tilde \Phi_{xyx} 
      \right), \nonumber\\
\fl  \zeta \tilde h^{(\mathrm{C})}_3 \doteq \tilde c_y^{0-} =& 
    \partial_\xi \left( -\tilde \Pi_{\st y} - \tilde \Pi_{xy} 
      - \tilde \Phi_{x\st y} - \tilde \Phi_{xxy} \right) \nonumber \\ & 
    + \rmi \left( -\half \tilde \Pi_{\st\st} + \half \tilde \Pi_{xx} 
      - \half \tilde \Pi_{yy} + \half \tilde \Pi_{zz}  
      \right. \\ & \qquad \left. 
      - \tilde \Phi_{y\st y} - \half \tilde \Phi_{xyy} 
      - \half \tilde \Phi_{x\st\st} + \half \tilde \Phi_{xxx} 
      + \half \tilde \Phi_{xzz} \right) , \nonumber\\
  \label{eq:flc0mbc4}
\fl  \zeta \tilde h^{(\mathrm{C})}_4 \doteq \tilde c_z^{0-} =& 
    \partial_\xi \left( - \tilde \Pi_{\st z} - \tilde \Pi_{xz} 
      - \tilde \Phi_{x\st z} - \tilde \Phi_{xxz}  \right) 
    + \rmi \left( - \tilde \Pi_{yz} - \tilde \Phi_{y\st z} 
      - \tilde \Phi_{xyz} \right) .
\end{eqnarray}
In the $\beta < 0$ case, we have the additional constraint-preserving
boundary conditions \eref{eq:c4bc},
\begin{eqnarray}
  \label{eq:flc4bc1}
  \zeta \tilde h^{(\mathrm{C})}_{5,ab} \doteq \tilde c^4_{xyab} = 
    \partial_\xi \tilde \Phi_{yab} - \rmi \tilde \Phi_{xab},\\
  \label{eq:flc4bc2}
  \zeta \tilde h^{(\mathrm{C})}_{6,ab} \doteq \tilde c^4_{xzab} = 
    \partial_\xi \tilde \Phi_{zab} .
\end{eqnarray}
Although it has to vanish for a solution that satisfies the constraints,
we have added boundary data $\bi h^{(\mathrm{C})}$ to all the 
constraint-preserving boundary conditions in order to account for the 
inevitable numerical truncation error.

The physical boundary conditions \eref{eq:physbc} become
\begin{eqnarray}
  \label{eq:flphysbc1}
\fl   \zeta \tilde h^{(\mathrm{P})}_{1} \doteq \tilde w^-_{yy} =& 
    \partial_\xi \left(  - \half \tilde \Pi_{yy} + \half \Pi_{zz} 
      - \half \tilde \Phi_{xyy} + \half \tilde \Phi_{xzz}
      \right) \\ &
    + \rmi \left( \half \tilde \Pi_{yx} + \tilde \Phi_{xxy} + \tilde \Phi_{zzy} 
      - \half \tilde \Phi_{yxx} - \half \tilde \Phi_{xy\st} 
      + \half \tilde \Phi_{yx\st} \right) ,\nonumber\\
  \label{eq:flphysbc2}
\fl  \zeta \tilde h^{(\mathrm{P})}_{2} \doteq \tilde w^-_{yz} =&
    \partial_\xi \left( - \tilde \Pi_{yz} -\tilde \Phi_{xyz}  \right) 
      \\&
    + \rmi \left( \half \tilde \Pi_{zx} + \tilde \Phi_{xxz} 
      + \half \tilde \Phi_{zzz} - \half \tilde \Phi_{zxx} 
      - \half \tilde \Phi_{zyy} 
      - \half \tilde \Phi_{xz\st} + \half \tilde \Phi_{zx\st} \right).\nonumber
\end{eqnarray}
Note that the terms involving a normal derivative of $u^0_{ab} \equiv
\psi_{ab}$ in \eref{eq:weylexpr} do not contribute because the
Fourier-Laplace transform of $\psi_{ab}$ vanishes as a consequence of
the evolution equations.

Finally, the gauge boundary conditions \eref{eq:gaugebc1} are
\begin{eqnarray}
  \label{eq:flgaugebc1}
  \tilde h^{(\mathrm{G})}_{1} \doteq
     \tilde \Pi_{\st\st} - \tilde \Pi_{xt} + \tilde \Phi_{x\st\st}
     - \tilde \Phi_{xx\st}, \\
  \tilde h^{(\mathrm{G})}_{2} \doteq
     \tilde \Pi_{\st x} - \tilde \Pi_{xx} + \tilde \Phi_{x\st x}
     - \tilde \Phi_{xxx}, \\
  \tilde h^{(\mathrm{G})}_{3} \doteq
     \tilde \Pi_{\st y} - \tilde \Pi_{xy} + \tilde \Phi_{x\st y}
     - \tilde \Phi_{xxy}, \\
  \label{eq:flgaugebc4}
  \tilde h^{(\mathrm{G})}_{4} \doteq
     \tilde \Pi_{\st z} - \tilde \Pi_{xz} + \tilde \Phi_{x\st z}
     - \tilde \Phi_{xxz} .
\end{eqnarray}

The next step in the calculation is to insert the general $L^2$ solutions
into the Fourier-Laplace transforms of the boundary conditions. 
We obtain a system of linear equations
\begin{equation}
  C(\zeta) \bsigma = \tilde \bi h .
\end{equation}
We need to evaluate the determinant of the matrix $C$.
The resulting expressions are rather lengthy and some computer algebra 
is helpful here. A programme (available from the author upon request)
has been written in the computer algebra language REDUCE \cite{REDUCE} 
in order to obtain the following results.
We discuss the cases $\beta \geqslant 0$ and $\beta < 0$ separately.

Suppose first that $\beta \geqslant 0$. In this case the general 
solution is \eref{eq:posbetasoln} and the boundary conditions are 
(\ref{eq:flc0mbc1}--\ref{eq:flc0mbc4},
\ref{eq:flphysbc1}--\ref{eq:flgaugebc4}).
We find
\begin{equation}
  \det C = \frac{-[\zeta - (1 + \beta) \lambda_1^-]^{16}}{8 \zeta^6 K_1^{10}}.
\end{equation}
The function $\zeta \rightarrow \zeta - (1 + \beta) \lambda_1^-$ maps
the right half of the complex plane to the right half of the complex
plane minus a circle of radius $[(1 + \beta)/(1 - \beta)]^{1/2}$.
Hence $\det C$ has no zeros $\zeta$ with $\Re(\zeta) > 0$ and is
bounded away from zero even as $\Re(\zeta) \rightarrow 0$. 
In the limit $|\zeta| \rightarrow \infty$, we have 
$|\lambda_1^-| \sim |\zeta|$, $|K_1| \sim |\zeta|$
and hence $\det C$ is bounded away from zero, too. We conclude that
both the determinant condition and the Kreiss condition are satisfied.

Next, we consider the case $\beta < 0$. 
Suppose first that $\zeta \neq -\beta$ so that the general decaying
solution is \eref{eq:negbetasoln}. Plugging this into the boundary
conditions (\ref{eq:flc0mbc1}--\ref{eq:flgaugebc4}) yields
\begin{equation}
  \det C = \frac{-[\zeta - (1 + \beta) \lambda_1^-]^{16} (\zeta^2 - \beta^2)^{10}}
  {8 \zeta^{16} \beta^{20} K_1^{10} K_2^{10}}.
\end{equation}
By the same argument as above, $\det C$ has no zeros $\zeta \neq -\beta$ 
with $\Re(\zeta) \geqslant 0$.
In the degenerate case $\zeta = -\beta$, we have to use 
the solution \eref{eq:specnegbetasoln} instead and find
\begin{equation}
  \det C = \frac{(K_1 + K_2)^{10}}{8 \beta^{26} K_1^{10} K_2^{20}} \neq 0.
\end{equation}

We conclude that for arbitrary shift at the boundary (with subluminal
normal component), 
our initial-boundary value problem satisfies the determinant condition
and the Kreiss condition in the high-frequency limit.
In particular, there are no strong instabilities.

It is easy to see that this result is unchanged if the physical
boundary conditions \eref{eq:physbc} are replaced with the improved
conditions \eref{eq:improvedphysbc} of \cite{BuchmanSarbach}.
In the frozen-coefficient approximation used in this section, 
they correspond to successive applications of the operator 
$(t^a + n^a) \partial_a =  \hat \partial_t - (1 + \beta) \partial_x$ 
to the boundary conditions \eref{eq:physbc}.
Each produces an additional factor of $\zeta - (1 + \beta) \lambda_1^-$ 
in the expressions for $\det C$, and as argued above, this factor is 
bounded away from zero.

Finally, we ask what happens if we replace the gauge boundary
conditions \eref{eq:gaugebc} with the alternative conditions
\eref{eq:altgaugebc}. In this case, we obtain
\begin{equation} \fl
  \det C = \cases{
    \frac{-[\zeta - (1 + \beta) \lambda_1^-]^{8} 
    (\zeta - \beta \lambda_1^-)^8}{8 \zeta^6 K_1^{10}}
    & if $\beta \geqslant 0$, \medskip\\ 
    \frac{-[\zeta - (1 + \beta) \lambda_1^-]^{8} 
    (\zeta - \beta \lambda_1^-)^8 (\zeta^2 - \beta^2)^{10}}
    {8 \zeta^{16} \beta^{20} K_1^{10} K_2^{10}}
    & if $\beta < 0$, }
\end{equation}
so that $\det C$ always has a zero
\begin{equation}
  \label{eq:badzero}
  \zeta = -\beta.
\end{equation}
Both the determinant condition and the Kreiss condition
are satisfied if the shift points towards the interior at the boundary
($\beta > 0$). However if the shift points towards the exterior 
($\beta < 0$), the IBVP is ill-posed. For tangential shift ($\beta = 0$), 
the determinant condition is satisfied (and hence there are no strong
instabilities) but the Kreiss condition is violated ($\zeta = 0$ is a
generalized eigenvalue). What happens in this case will be
investigated in the following section.


\section{Weak instabilities}
\label{sec:weakinstab}

As we have seen in the previous section, the determinant condition 
\eref{eq:gendetcond} is sufficient to detect strong instabilities 
of the form \eref{eq:genflansatz}. The question remains whether a given
IBVP admits ill-posed modes with ``milder'' than exponential time
dependence, so-called \emph{weak instabilities}. 
Examples of such instabilities with linear time dependence have been found in
a formulation of Maxwell's equations in a gauge with vanishing 
electrostatic potential \cite{ReulaSarbach} and in an Einstein-Christoffel 
formulation of general relativity \cite{SarbachTiglio}.
 
In this section, we develop a technique that can be used to 
systematically search for weak instabilities with \emph{polynomial}
time dependence.
The criterion turns out to be closely related to the Kreiss condition.
We then apply our method to the generalized harmonic system and show that
it does not suffer from such instabilities. 
Furthermore, we show that if we use the alternative gauge boundary
conditions \eref{eq:altgaugebc} instead, a weak instability does exist,
and we construct it explicitly.

\subsection{General theory}
\label{sec:genweak}

We consider again the general IBVP (\ref{eq:geneveqs}--\ref{eq:genbcs}).
As before, we Fourier-transform with respect to the coordinates $x^A$ 
tangential to the boundary. However, instead of assuming an exponential time
dependence as in the Laplace transform \eref{eq:genflansatz}, we now look
for modes whose time dependence is given by a polynomial of order 
$p \geqslant 1$, i.e., we make the ansatz
\begin{equation}
  \label{eq:wansatz}
  \bi u(t, x^i) = \exp(\rmi \omega \hat \omega_A x^A) \sum_{\nu = 0}^p 
  \tilde \bi u^{(\nu)} (\omega x^1) \frac{(\omega t)^\nu}{\nu !},
\end{equation}
where $\omega > 0$, $\hat \omega_A \hat \omega^A = 1$, 
and $\tilde \bi u^{(p)} \neq \bf 0$.
Substituting this into the evolution equations \eref{eq:geneveqs} and
evaluating order by order in $t$, we find (setting $\xi \equiv \omega x^1$)
\begin{eqnarray}
  \label{eq:wode1}
  A^1 \partial_\xi \tilde \bi u^{(p)} + \rmi \hat \omega_A A^A 
  \tilde \bi u^{(p)} = 0, \\
  \label{eq:wode2}
  A^1 \partial_\xi \tilde \bi u^{(p-1)} + \rmi \hat \omega_A A^A 
  \tilde \bi u^{(p-1)} = \tilde \bi u^{{(p)}} , \\
  \label{eq:wode3}
  A^1 \partial_\xi \tilde \bi u^{(\nu)} + \rmi \hat \omega_A A^A 
  \tilde \bi u^{(\nu)} = \tilde \bi u^{{(\nu+1)}} , \qquad
  0 \leqslant \nu \leqslant p-2.
\end{eqnarray} 

Let us also expand the boundary data $\bi h$ in a similar fashion as
in \eref{eq:wansatz}, up to order $p-1$ for a reason that will become
clear shortly,
\begin{equation}
  \bi h(t, x^A) = \exp(\rmi \omega \hat \omega_A x^A)
     \sum_{\nu = 0}^{p-1} \tilde \bi h^{(\nu)} \frac{(\omega t)^\nu}{\nu!}.
\end{equation}
The boundary conditions \eref{eq:genbcs} then read
\begin{eqnarray}
  \label{eq:wbc1}
  S^1 \partial_\xi \tilde \bi u^{(p)} + \rmi \hat \omega_A S^A 
  \tilde \bi u^{(p)} \doteq 0, \\
  \label{eq:wbc2}
  S^1 \partial_\xi \tilde \bi u^{(p-1)} + \rmi \hat \omega_A S^A 
  \tilde \bi u^{(p-1)} \doteq 0, \\
  \label{eq:wbc3}
  S^1 \partial_\xi \tilde \bi u^{(\nu)} + \rmi \hat \omega_A S^A 
  \tilde \bi u^{(\nu)} \doteq \tilde \bi h^{(\nu+1)}, \qquad
  0 \leqslant \nu \leqslant p-2. 
\end{eqnarray}

Suppose that a solution $\tilde \bi u = (\tilde \bi u^{(0)},
\tilde \bi u^{(1)}, \ldots, \tilde \bi u^{(p)})$ satisfying equations 
(\ref{eq:wode1}--\ref{eq:wode3}) and (\ref{eq:wbc1}--\ref{eq:wbc3}) exists.
Then the solution at times $t > 0$ is of order $\Or (\omega^p)$ whereas
the initial data is $\Or (\omega^0)$ and (the time integral of) the
boundary data is $\Or (\omega^{p-1})$.
Hence the solution cannot be estimated in terms of the initial and 
boundary data as $\omega \rightarrow \infty$ and the IBVP is ill-posed. 
We may actually argue that it is already ill-posed if a (nonzero) solution of
(\ref{eq:wode1}--\ref{eq:wode3}) exists that obeys the top two boundary
conditions (\ref{eq:wbc1}--\ref{eq:wbc2}), 
for we can always choose the boundary data $\tilde \bi h$ such that the 
remaining boundary conditions \eref{eq:wbc3} are satisfied. 

Let us look more closely at equations \eref{eq:wode1} and \eref{eq:wbc1}:
these are identical with equations \eref{eq:genflt} and 
\eref{eq:genbcflt} in section \ref{sec:genfl} for $s = 0$.
We conclude that if a weak instability exists, $s = 0$ is a
generalized eigenvalue. (Strictly speaking, one has to consider the
Fourier-Laplace solutions for $\Re(s) > 0$ and take the limit 
$s \rightarrow 0$. We assume here that the solutions are continuous at 
$s = 0$, which is the case in all the examples we discuss.)
Therefore, \emph{if the Kreiss condition is satisfied, there are no
  weak instabilities of any polynomial order $p$.}

On the other hand, $s = 0$ being a generalized eigenvalue does not 
automatically imply that there is a weak instability. One still has to 
solve the coupled ODE system (\ref{eq:wode1}--\ref{eq:wode3}) 
and impose the boundary conditions (\ref{eq:wbc1}--\ref{eq:wbc2})
on its solution.

In \cite{ReulaSarbach} it is claimed that (for a certain choice of
parameters) the Maxwell system considered there admits a weak
instability (with $p = 1$) even though the Kreiss condition is satisfied.
This is obviously a contradiction to the result that we have just derived.
However, the eigenvectors $\bi w_j$ appearing in the general $L^2$
solution in \cite[eq.(51)]{ReulaSarbach} are normalized in such a way
that they become infinite as $s \rightarrow 0$, and the estimate 
\eref{eq:genkreisscond1} does not obtain.
If the correct normalization is used, one finds that $s = 0$ is indeed
a generalized eigenvalue.

\subsection{Application to the GH system}

To see how the method outlined above works in practice, we apply it to the
generalized harmonic Einstein equations in the high-frequency limit
(equations (\ref{eq:dtpsi0}--\ref{eq:dtphi0})).

First, we take the gauge boundary conditions to be \eref{eq:gaugebc}.
We have already seen in section \ref{sec:ghfl} that the
Kreiss condition is satisfied, and hence there are no weak
instabilities with polynomial time dependence.
(One can also carry out the explicit calculation below, obtaining the
same result.)

Next, we consider the alternative gauge boundary conditions
\eref{eq:altgaugebc}, and we take the shift to be tangential at the
boundary ($\beta = 0$). As discussed in section \ref{sec:ghfl}, the Kreiss
condition is violated in this case, with $s = 0$ being a
generalized eigenvalue. This indicates that there might be a weak instability.
Let us search for one with $p = 1$ and homogeneous boundary data.
Equation \eref{eq:wode1} implies
\begin{eqnarray}
  \label{eq:w1a}
  -\partial_\xi \tilde \Phi_{xab}^{(1)} - \rmi \tilde \Phi_{yab}^{(1)} = 0, \\
  \label{eq:w1c}
  -\rmi \tilde \Pi_{ab}^{(1)} = 0,
\end{eqnarray}
and \eref{eq:wode2} reads
\begin{eqnarray}
  \label{eq:w0a}
  -\partial_\xi \tilde \Phi_{xab}^{(0)} - \rmi \tilde \Phi_{yab}^{(0)} 
    = \tilde \Pi_{ab}^{(1)}, \\
  \label{eq:w0b}
  - \partial_\xi \tilde \Pi_{ab}^{(0)} = \tilde \Phi_{xab}^{(1)}, \\
  \label{eq:w0c}
  - \rmi \Pi_{ab}^{(0)} = \tilde \Phi_{yab}^{(1)}, \\
  0 = \tilde \Phi_{zab}^{(1)}. 
\end{eqnarray}
Substituting (\ref{eq:w0b},\ref{eq:w0c}) into \eref{eq:w1a}, we obtain
\begin{equation}
  \partial_\xi^2 \tilde \Pi_{ab}^{(0)} = \tilde \Pi_{ab}^{(0)}
  \quad \Rightarrow \quad
  \tilde \Pi_{ab}^{(0)} = \sigma_{1ab} e^{-\xi}.
\end{equation}
Assuming that $\C_{iab} = 0$ initially, i.e.,
\begin{equation}
  \tilde \Phi_{xab}^{(0)} = \omega \partial_\xi \tilde \psi_{ab}^{(0)}, \qquad
  \tilde \Phi_{yab}^{(0)} = \rmi \omega \tilde \psi_{ab}^{(0)}, \qquad
  \tilde \Phi_{zab}^{(0)} = 0,
\end{equation}
equation \eref{eq:w0a} together with \eref{eq:w1c} implies similarly
\begin{equation}
  \partial_\xi^2 \tilde \psi_{ab}^{(0)} = \tilde \psi_{ab}^{(0)}
  \quad \Rightarrow \quad
  \tilde \psi_{ab}^{(0)} \equiv \omega^{-1} \sigma_{2ab} e^{-\xi}.
\end{equation}
Hence the most general $L^2$ solution (consistent with $\C_{iab} = 0$) is
of the form
\begin{eqnarray}
  \left( \tilde \Pi_{ab}^{(0)}, \tilde \Phi_{xab}^{(0)}, \tilde \Phi_{yab}^{(0)}, 
    \tilde \Phi_{zab}^{(0)} \right) &=& 
  \rme^{-\xi} ( \sigma_{1ab}, -\sigma_{2ab}, \rmi \sigma_{2ab}, 0 ),\\
  \left( \tilde \Pi_{ab}^{(1)}, \tilde \Phi_{xab}^{(1)}, \tilde \Phi_{yab}^{(1)}, 
    \tilde \Phi_{zab}^{(1)} \right) &=& 
  \rme^{-\xi} ( 0, \sigma_{1ab}, -\rmi \sigma_{1ab}, 0 ).
\end{eqnarray}
Only the mode associated with $\sigma_{1ab}$ corresponds to a solution
with $\tilde \bi u^{(1)} \neq \bf 0$. 
Substituting this mode into the Fourier transform of the homogeneous
boundary conditions 
and evaluating order by order in $t$, one obtains a $20\times 10$
linear system of equations for the constants $\sigma_{1ab}$. 
The only nontrivial solution is given by
\begin{equation}
  \sigma_{1xx} = 1, \, \sigma_{1xy} = -\rmi, \, \sigma_{1yy} = -1, \quad 
  \textrm{all other } \, \sigma_{kab} = 0.
\end{equation}
It corresponds to setting
\begin{eqnarray}
  \Pi_{ij} = f_{,ij}, \quad \Pi_{0a} = 0, \nonumber\\
  \Phi_{kij} = -t f_{,kij}, \quad \Phi_{k0a} = 0, \\
  f \equiv \rme^{\omega(-x + \rmi y)}, \nonumber
\end{eqnarray}
which is of the same type as the weak instabilities found 
in \cite{ReulaSarbach, SarbachTiglio}.

A similar calculation shows that there are no weak instabilities of
order $p \geqslant 2$ for homogeneous boundary data.
However, as explained in section \ref{sec:genweak}, this is not true 
for inhomogeneous boundary data.


\section{Robust stability  tests}
\label{sec:num}

In this section, we perform a series of \emph{robust stability tests} 
on our numerical implementation of the GH system. This test involves 
adding small random perturbations to a known exact solution. 
Thus all possible frequencies at a given resolution are excited. 
By increasing the resolution and looking at the growth
rate of the numerical solution, this test is very effective in
spotting ill-posed modes.
The robust stability test was suggested in the context of the 
Cauchy problem by the ``Apples with Apples'' collaboration  
(\cite{AwA}, see also \cite{Boyle} for an application to the spectral 
evolution code used for this work). 
Previous studies in which a version of the test was used in relation
to stability of the IBVP include \cite{SzilagyiGomezBishopWinicour,
SzilagyiSchmidtWinicour,SzilagyiWinicour,SarbachTiglio,BonaCPBC}.

For most of the tests, the background solution is taken to be 
flat space with a constant shift $N^i$,
\begin{equation}
  \label{eq:flatshift}
  \rmd s^2 = -\rmd t^2 + \delta_{ij} (\rmd x^i + N^i \rmd t)
    (\rmd x^j + N^j \rmd t).
\end{equation}
We remark that this is the most general background solution in the
high-frequency limit, as discussed in section \ref{sec:ghfl}.
The spatial domain is taken to be of topology $T^2 \times \mathbb{R}$.
This domain is intended to be as simple as possible, avoiding additional 
complications due to sharp corners and edges.
We take $x,y,z \in [-0.5, 0.5]$. The boundary conditions discussed in
this paper are imposed at $x = \pm 0.5$, and periodic boundary
conditions are imposed in the $y$ and $z$ directions.

For the last test, we consider instead the Schwarzschild solution in
Kerr-Schild coordinates, 
\begin{equation}
  \label{eq:schwarzschild}
  \fl
  \rmd s^2 = - \left( 1 - \frac{2M}{r} \right) \rmd t^2 
    + \frac{4M}{r} \rmd r \, \rmd t
    + \left( 1 + \frac{2M}{r} \right) \rmd r^2 
    + r^2 \left( \rmd \theta^2 + \sin^2 \theta \rmd \phi^2 \right) .
\end{equation}
In this case, the spatial domain is taken to be a spherical shell 
extending from $r_\mathrm{min} = 1.8 M$ (just inside the event horizon) to 
$r_\mathrm{max} = 11.8 M$.
The boundary conditions discussed in this paper are imposed at
the outer boundary $r = r_\mathrm{max}$. No boundary conditions are imposed 
at the inner boundary $r = r_\mathrm{min}$.
See also \cite{Zinketal} for a spherically symmetric robust stability 
test on Schwarzschild spacetime in Painlev\'e-Gullstrand coordinates.

As initial data, we compute the variables $\psi_{ab}$, $\Pi_{ab}$ and
$\Phi_{iab}$ corresponding to the exact solutions (\ref{eq:flatshift}, 
\ref{eq:schwarzschild}) and add random noise of amplitude 
$\epsilon = 10^{-10}$ (in the flat-space case) or $\epsilon = 10^{-6}$ 
(in the Schwarzschild case) to them.
In addition, random noise of the same amplitude is added both to the
right-hand-sides of the expressions for the time derivatives at the
boundary \eref{eq:addingbcs} and to the right-hand-sides 
of the evolution equations. In this way, we probe the influence of nonzero
initial and source data as well, which could not be analyzed using the
methods of section \ref{sec:fl}.

For more physically realistic tests involving black hole spacetimes, 
we refer the reader to \cite{GH}. In particular, an ingoing gravitational 
wave was injected through the outer boundary (i.e., data $h^{(\mathrm{P})}$
were prescribed in equation \eref{eq:physbc}). The gravitational radiation 
scattered back by the black hole was extracted, correctly reproducing the
dominant quasi-normal mode oscillation.

\subsection{Numerical implementation}
\label{sec:numimpl}

We use pseudospectral methods as described for example in \cite{KidderBC}.

For the flat-space tests on $T^2 \times \mathbb{R}$, the numerical 
solution is expanded into Chebyshev polynomials in $x$ and into 
Fourier series in $y$ and $z$.
We use between $9$ and $27$ basis functions in the $x$ and $y$
directions (these are typical resolutions in realistic numerical 
relativity simulations using spectral methods) and $3$ basis functions 
in the $z$ direction. Reducing the effective dimension by one enables 
us to evolve up to final times of $t = 1000$ within a tolerable runtime of the 
code even for the higher resolutions. We have checked for moderate resolutions
that the behaviour is qualitatively unchanged in fully three-dimensional runs.
A fourth-order Runge-Kutta scheme is used for the time integration,
with a Courant factor of $\Delta t / \Delta x_\mathrm{min} = 1.5$,
where $x_\mathrm{min}$ is the minimum spacing between the
(in the Chebyshev case, non-uniform) pseudospectral collocation points.

For the Schwarzschild test, the numerical solution is expanded into
Chebyshev polynomials in $r$ and spherical harmonics in the angular
directions. Here we use between $15$ and $27$ radial and between $11$
and $19$ angular basis functions. The same time integration scheme
with the same Courant factor is used as in the flat-space case.
The highest-order spherical harmonics are filtered as described for
example in \cite{KidderBC}. No filtering is applied to the Chebyshev
and Fourier expansions.

The boundary conditions are implemented by prescribing the time
derivatives of the characteristic variables at the boundary \cite{Bjorhus}.
For the Neumann-like constraint-preserving boundary conditions 
(\ref{eq:c0mbc},\ref{eq:c4bc}) and physical boundary conditions 
\eref{eq:physbc}, we use the evolution equations for the incoming 
characteristic variables in order to treat normal derivatives for 
time derivatives,
\begin{eqnarray}
  d_t u^{1-}_{ab} &=& (N^n + N) d_n u^{1-}_{ab} +
  \textrm{tangential derivatives} ,\\
  d_t u^{2}_{Aab} &=& N^n d_n u^{2}_{Aab} + 
  \textrm{tangential derivatives} .
\end{eqnarray}
(The notation $d$ instead of $\partial$ indicates that the partial
derivative acts only on the fundamental variables, not on the
eigenvectors needed to form the characteristic variables \cite{GH}.)
For the gauge boundary conditions, we simply take a time derivative of
\eref{eq:gaugebc}.
In this way, we obtain expressions for $d_t u^2$,
$P^{(\mathrm{C})} d_t u^{1-}$, $P^{(\mathrm{P})} d_t u^{1-}$, and 
$P^{(\mathrm{G})} d_t u^{1-}$ (omitting the spacetime indices 
for simplicity). These are then combined to form
\begin{equation}
  \label{eq:addingbcs}
  d_t u^{1-} \doteq P^{(\mathrm{C})} d_t u^{1-} + P^{(\mathrm{P})} d_t u^{1-}
  + P^{(\mathrm{G})} d_t u^{1-}
\end{equation}
at the boundary (note that $P^{(\mathrm{C})}$, $P^{(\mathrm{P})}$ and 
$P^{(\mathrm{G})}$ are
mutually orthogonal projection operators adding up to the identity).
In order to implement the alternative gauge boundary conditions
\eref{eq:altgaugebc}, we first fill $P^{(\mathrm{C})} d_t u^{1-}$ and
$P^{(\mathrm{P})} d_t u^{1-}$ as before and then set
\begin{equation}
  P^{(\mathrm{G})} d_t u^{1-} \doteq Q P^{(\mathrm{C})} d_t u^{1-} 
  + (Q - P^{(\mathrm{G})}) d_t (u^{1+} + 2 \gamma_2 \psi), 
\end{equation}
where the operator $Q$ has the properties 
$P^{(\mathrm{C})} Q = P^{(\mathrm{P})} Q = 0$,
$P^{(\mathrm{G})} Q = Q$, $Q P^{(\mathrm{C})} = Q$, 
$Q P^{(\mathrm{P})} = Q P^{(\mathrm{G})} = 0$, and
$t \cdot (P^{(\mathrm{C})} + Q) = 0$. 
It then follows from \eref{eq:addingbcs} and \eref{eq:u1m} that 
$d_t (t \cdot \Pi) \doteq 0$ 
at the boundary, as desired. An explicit expression for $Q$ is given by
\begin{equation}
  Q_{ab}{}^{cd} = k_a k_b k^c k^d + 2 k_{(a} P_{b)}{}^{(c} k^{d)}
  - 2 k_{(a} l_{b)} k^c k^d .
\end{equation}

\subsection{Numerical results}

\begin{figure}[t]
  \begin{minipage}[t]{0.49\textwidth}
    \includegraphics[scale = .26]{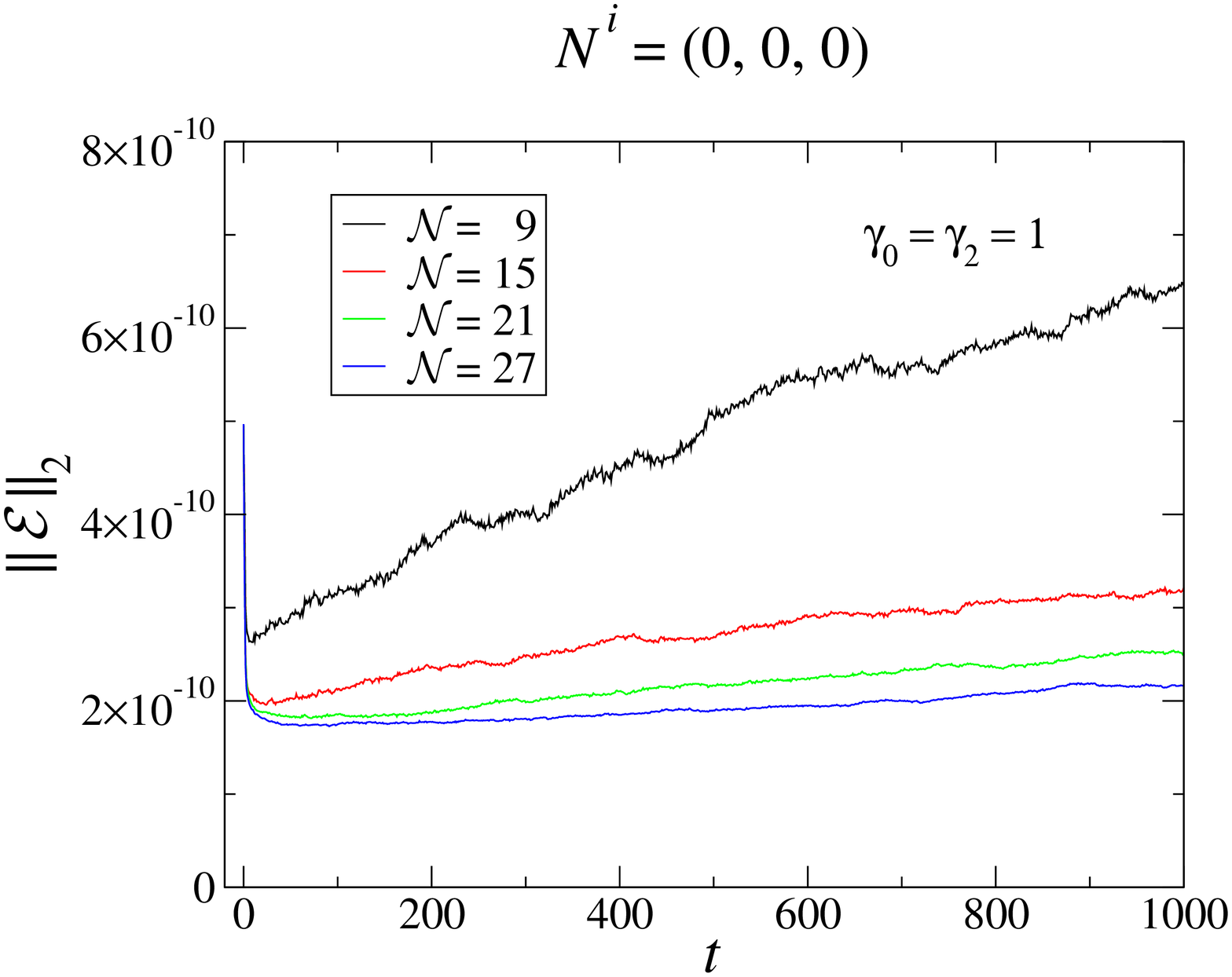}
  \end{minipage}
  \hfill  
  \begin{minipage}[t]{0.49\textwidth}
    \includegraphics[scale = .26]{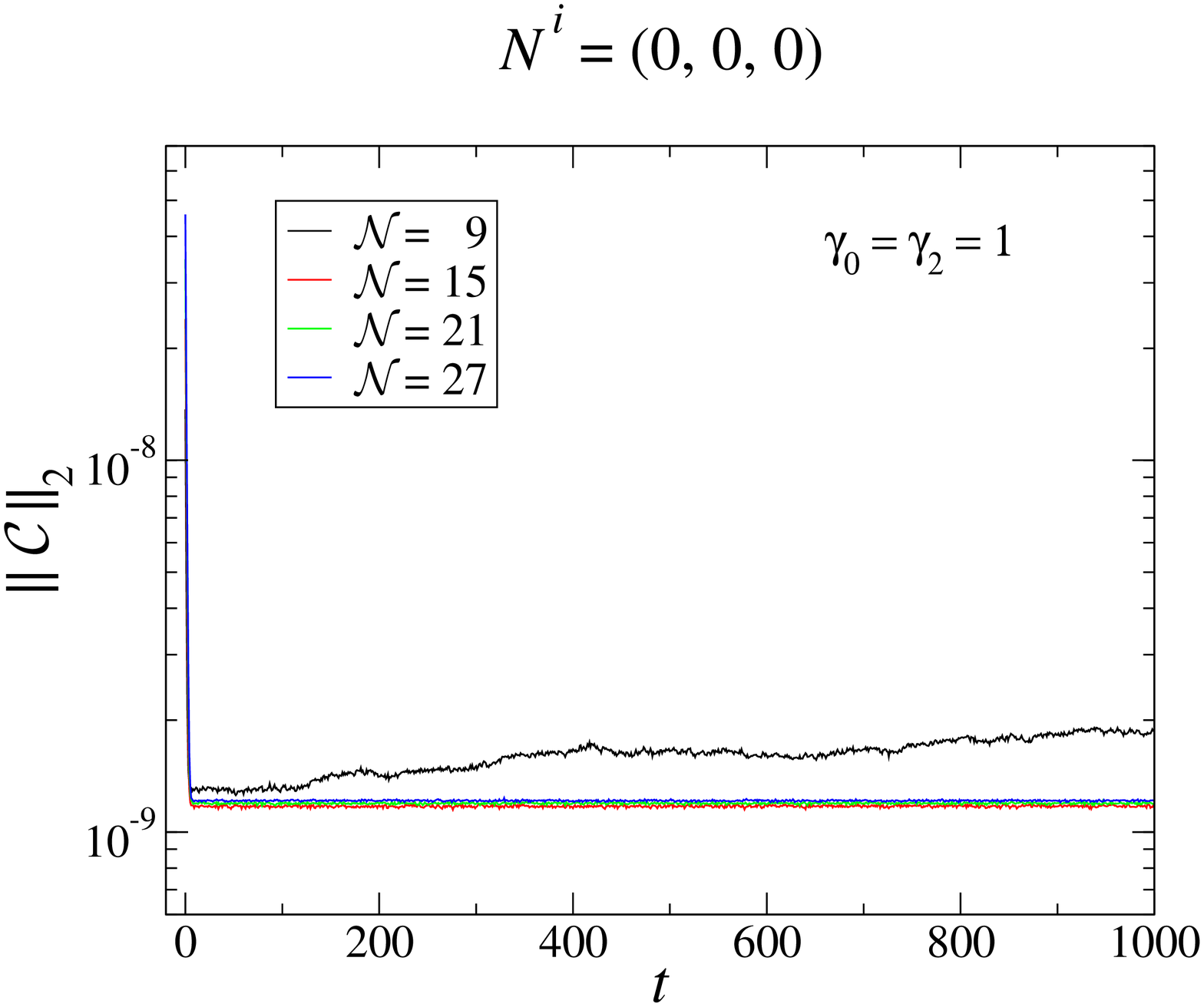}
  \end{minipage}
  \hfill  
  \begin{minipage}[t]{0.49\textwidth}
    \includegraphics[scale = .26]{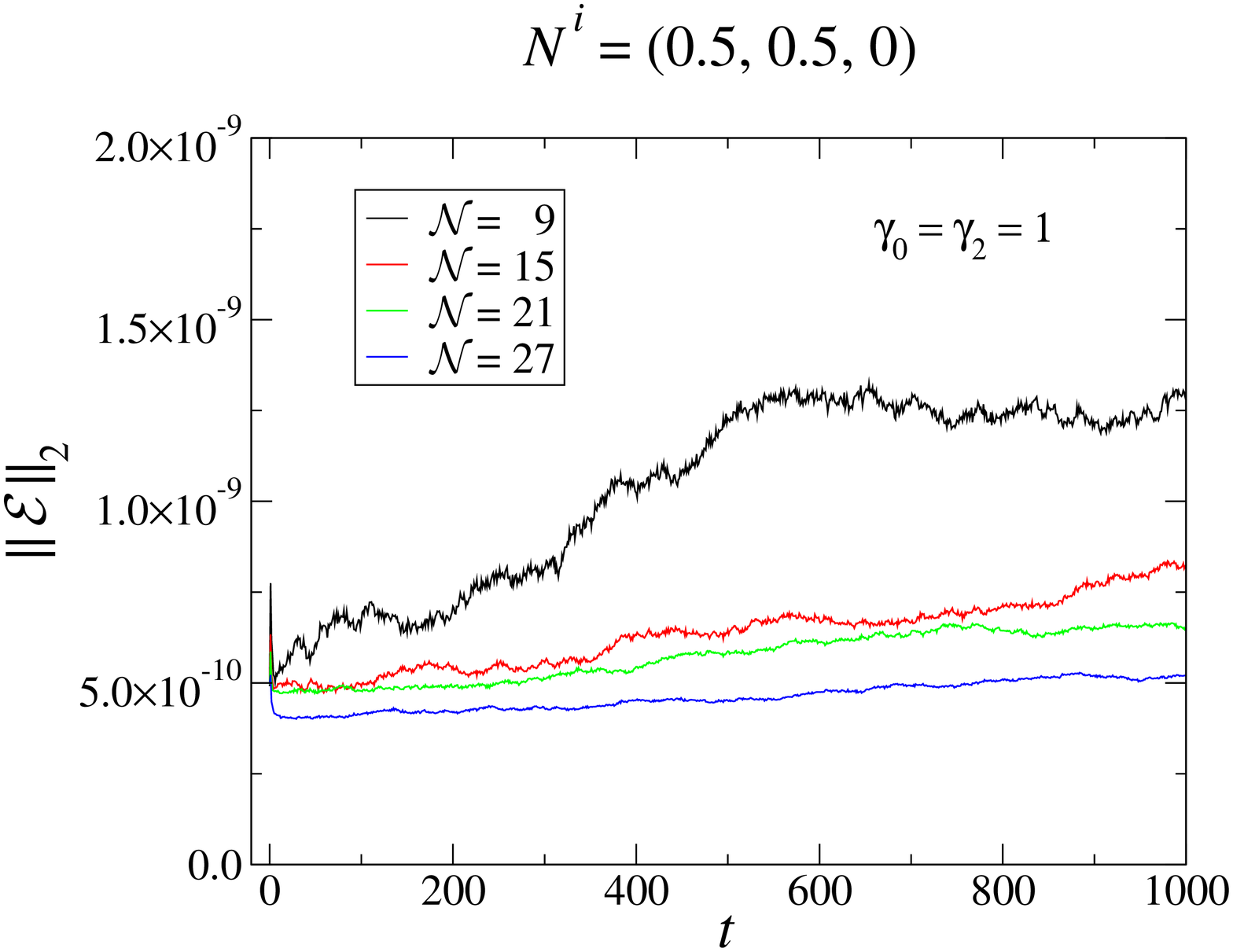}
  \end{minipage}
  \hfill  
  \begin{minipage}[t]{0.49\textwidth}
    \includegraphics[scale = .26]{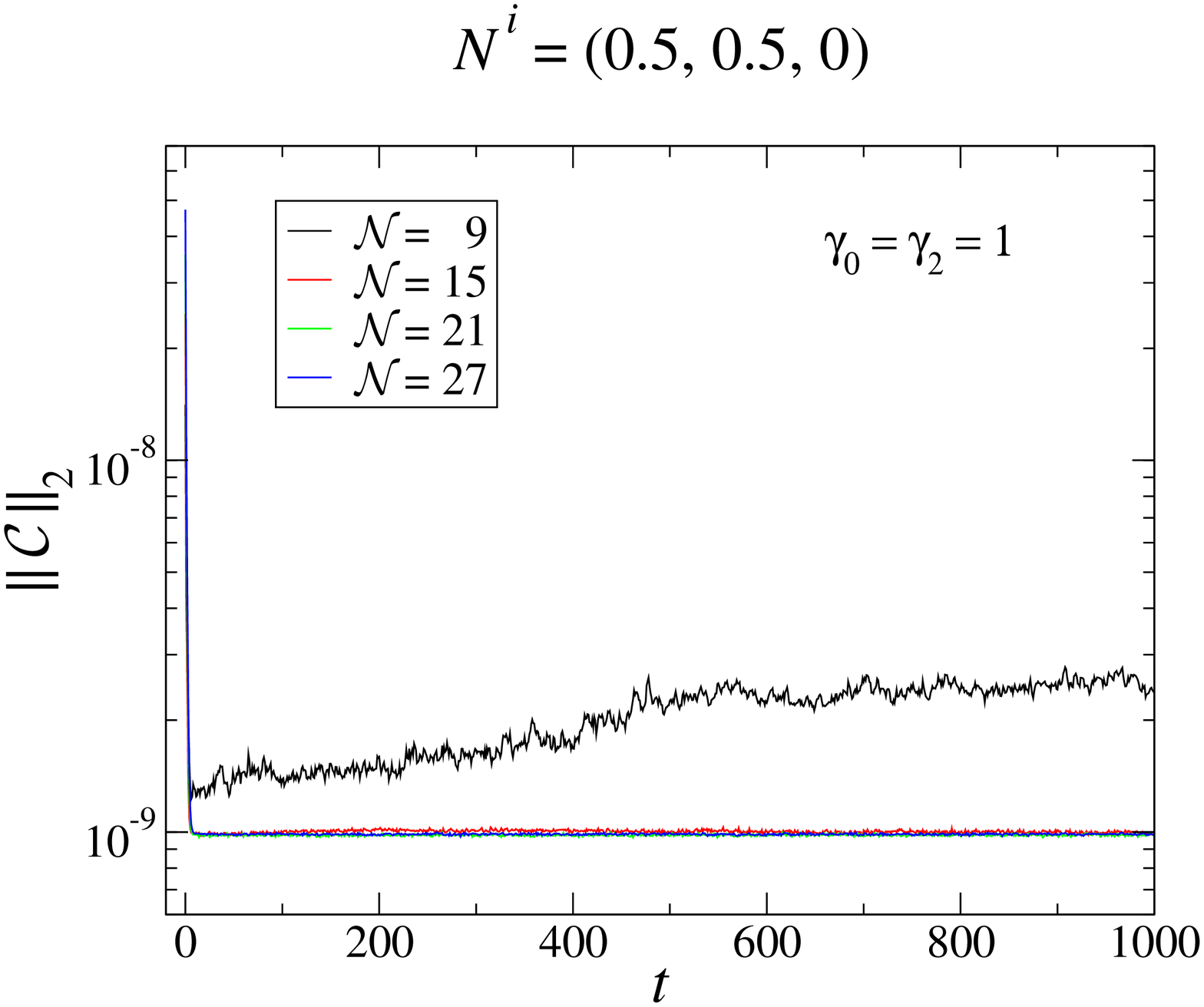}
  \end{minipage}
  \caption{\label{fig:robstabgood} 
  Robust stability test on the flat-space background \eref{eq:flatshift} 
  \emph{with} constraint-damping.
  $L^2$-norms of the error (left) and the constraints (right) 
  for two different shift vectors $N^i$ and for four different 
  resolutions $\N \equiv \N_x = \N_y$. }
\end{figure}

\begin{figure}[t]
  \begin{minipage}[t]{0.49\textwidth}
    \includegraphics[scale = .26]{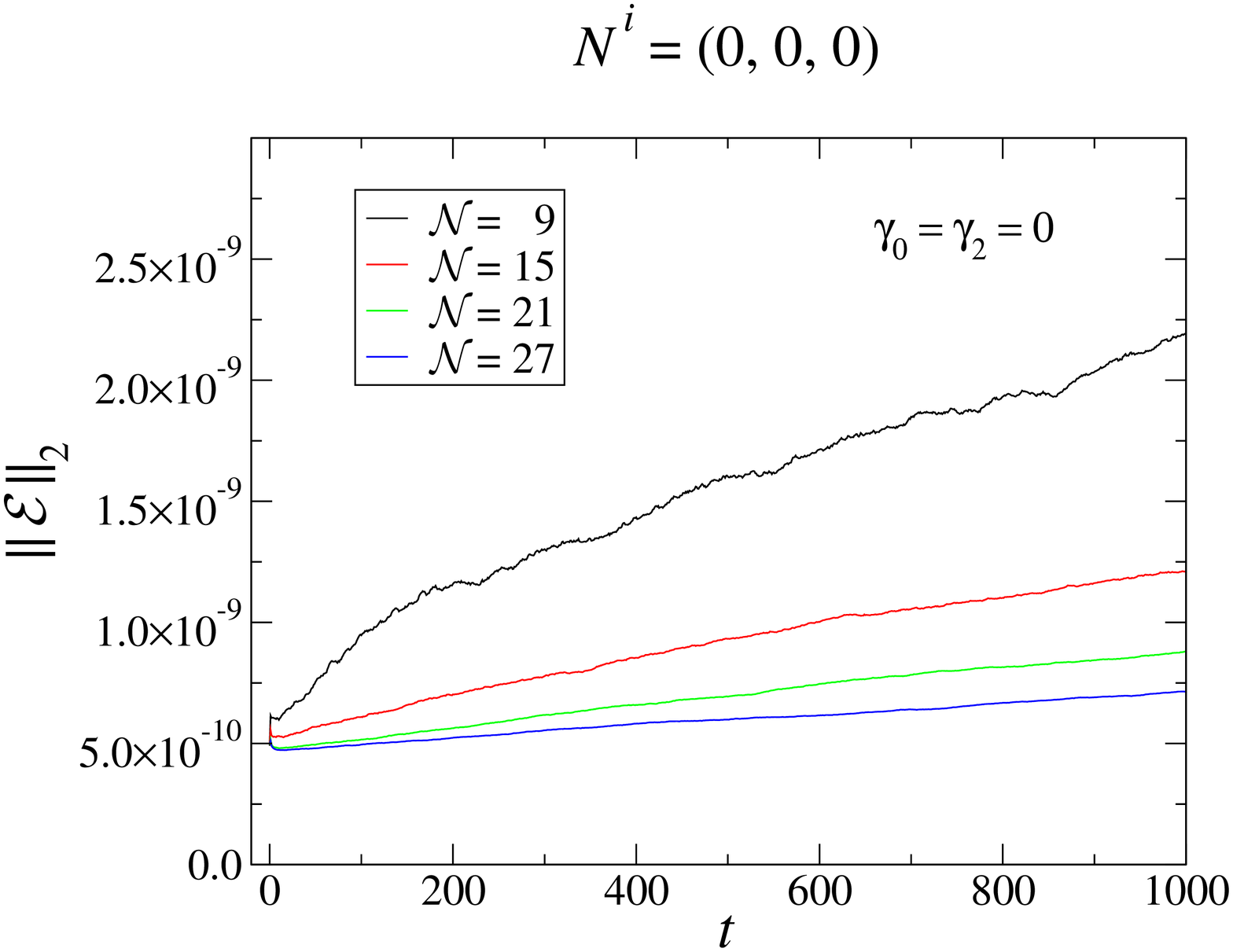}
  \end{minipage}
  \hfill  
  \begin{minipage}[t]{0.49\textwidth}
    \includegraphics[scale = .26]{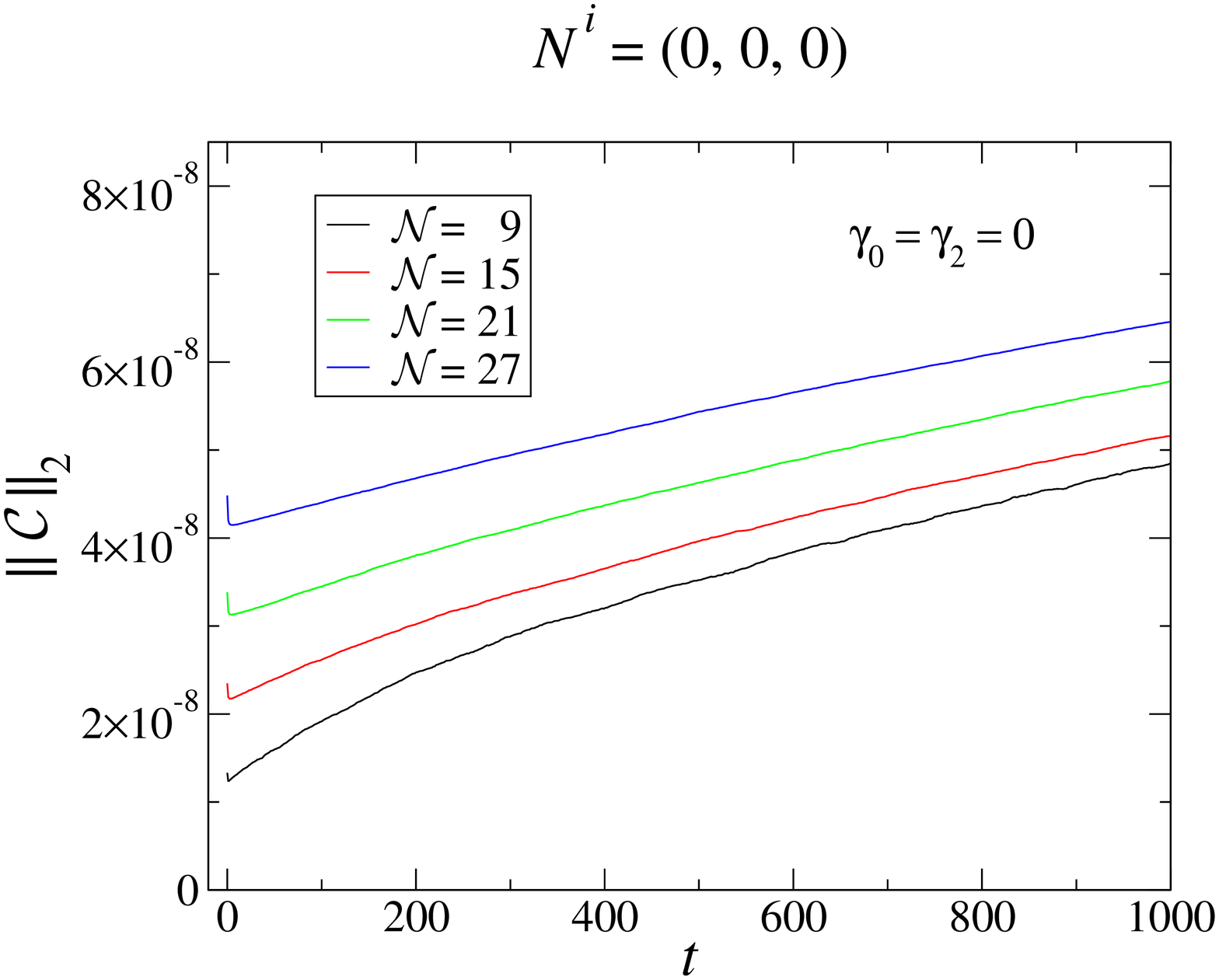}
  \end{minipage}
  \hfill  
  \begin{minipage}[t]{0.49\textwidth}
    \includegraphics[scale = .26]{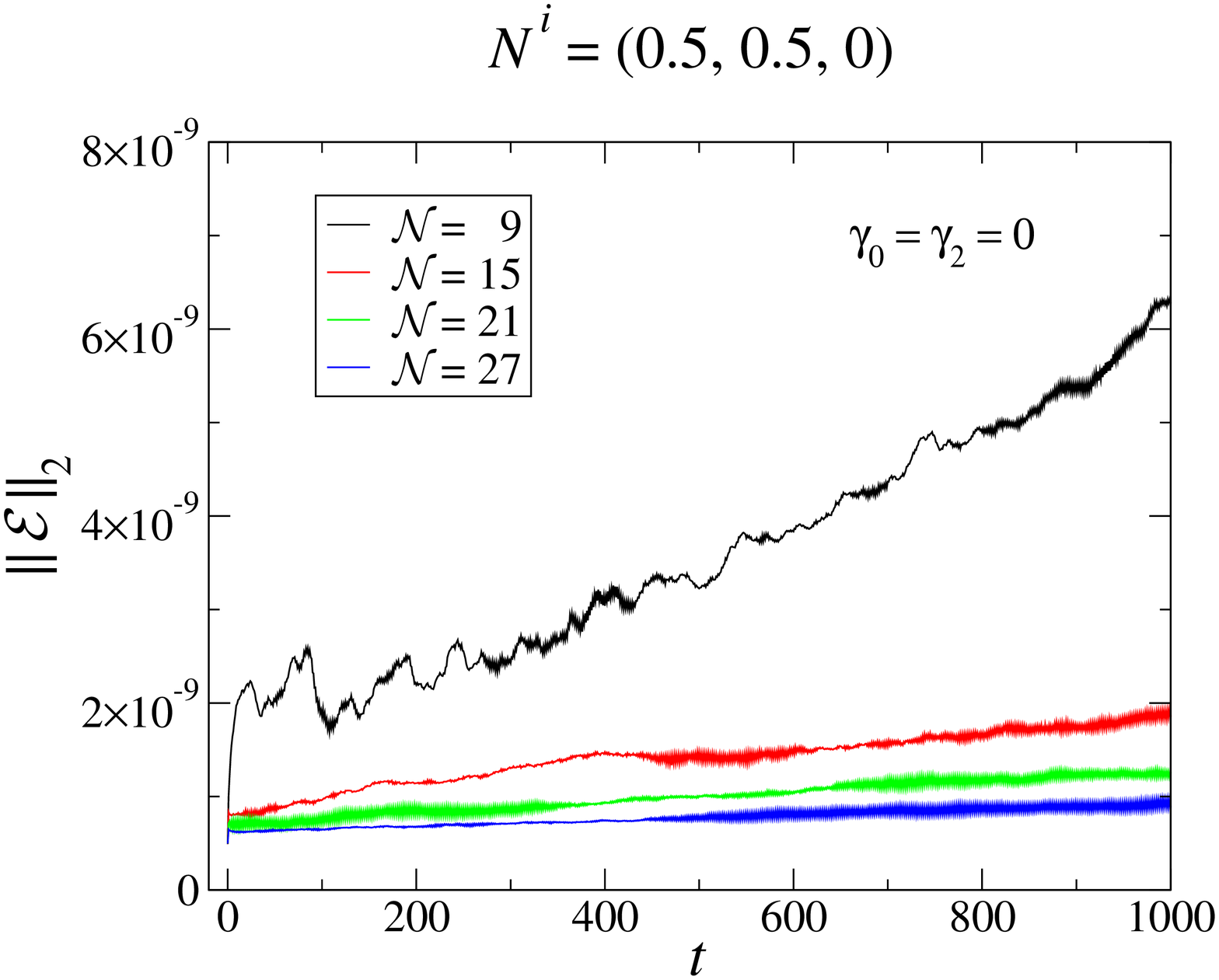}
  \end{minipage}
  \hfill  
  \begin{minipage}[t]{0.49\textwidth}
    \includegraphics[scale = .26]{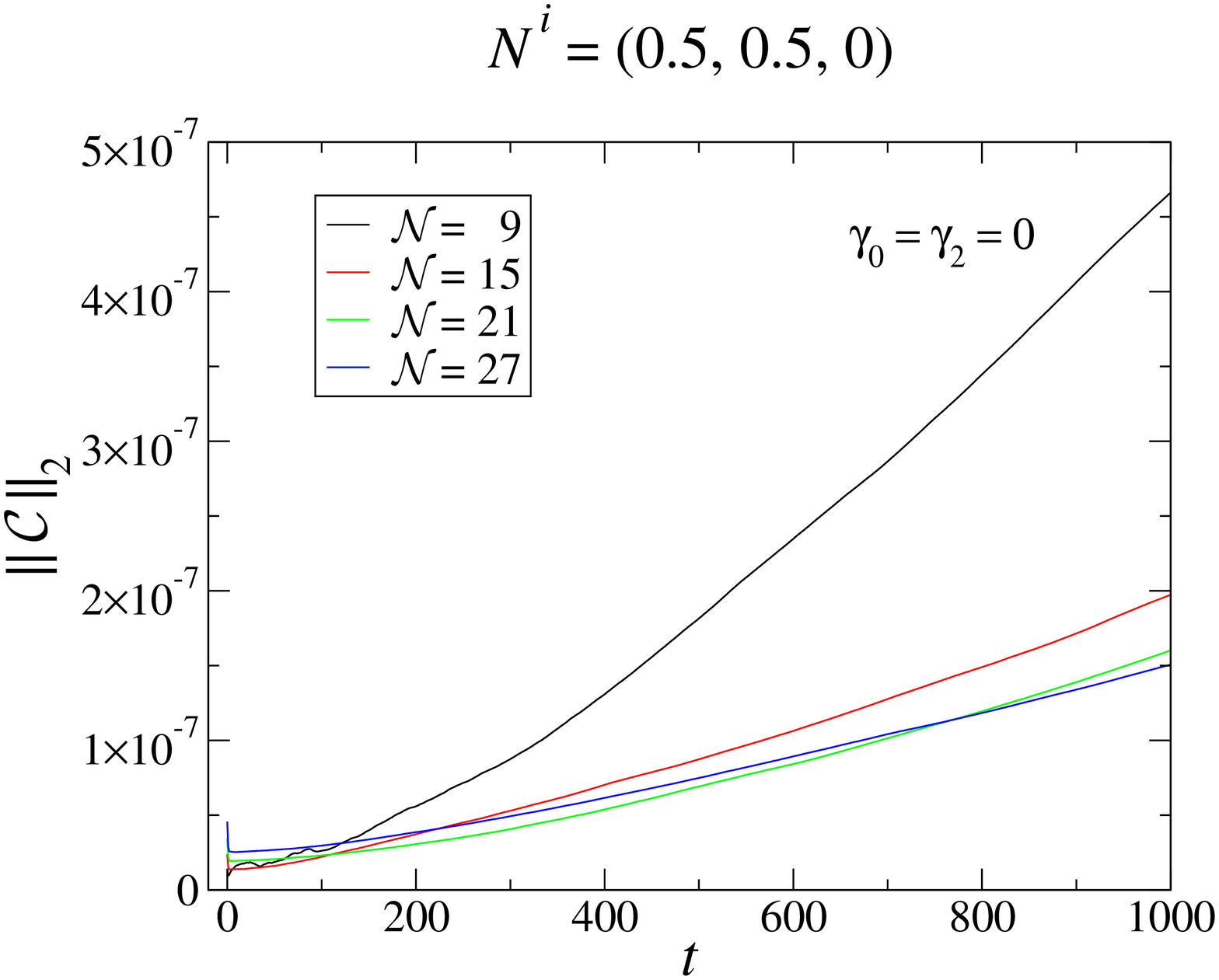}
  \end{minipage}
  \caption{\label{fig:robstabgoodnodamp} 
  Robust stability test on the flat-space background \eref{eq:flatshift}
  \emph{without} constraint damping. 
  $L^2$-norms of the error (left) and the constraints (right) 
  for two different shift vectors $N^i$ and for four different 
  resolutions $\N \equiv \N_x = \N_y$.}
\end{figure}

\begin{figure}[t]
  \begin{minipage}[t]{0.49\textwidth}
    \includegraphics[scale = .26]{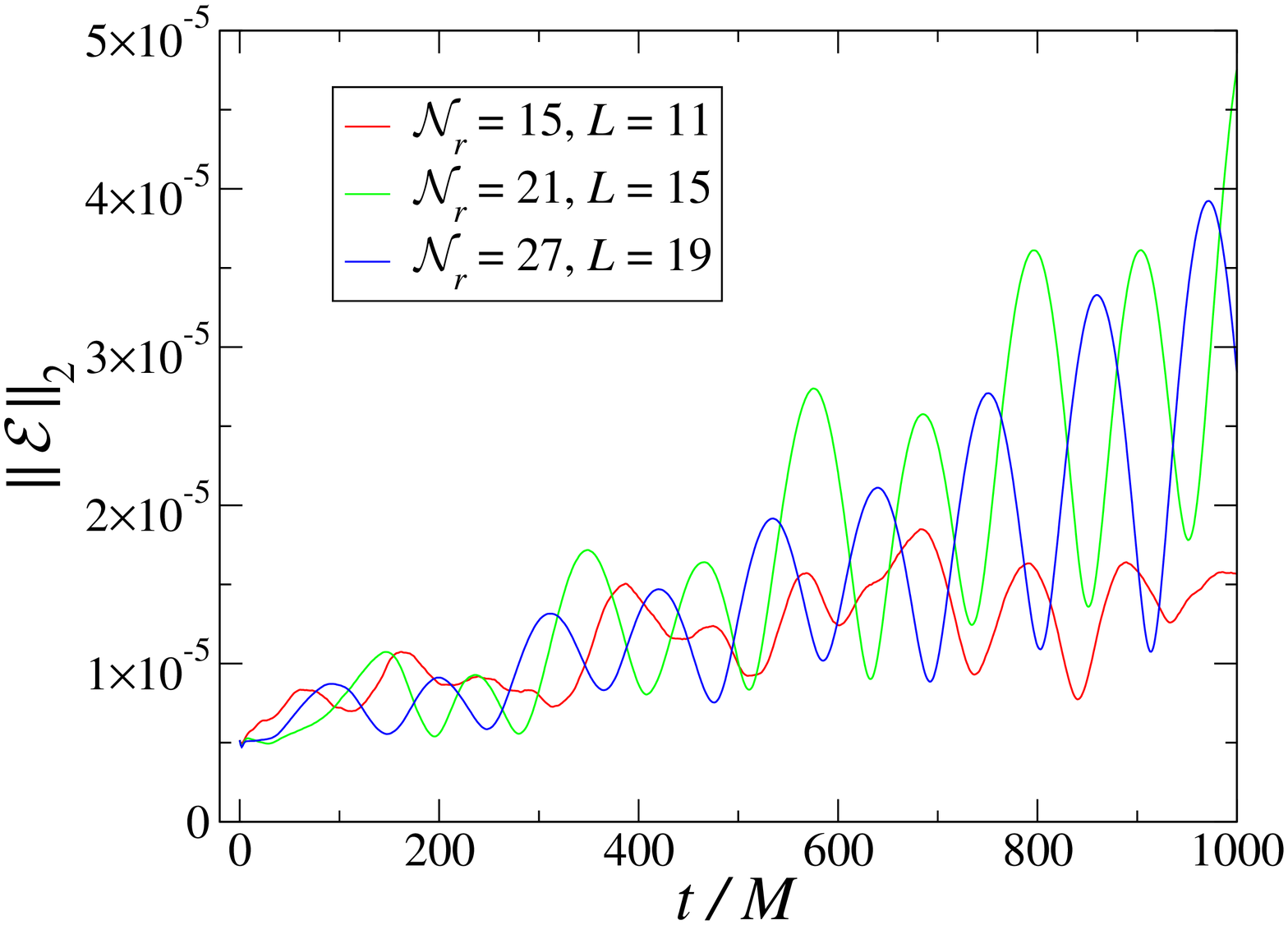}
  \end{minipage}
  \hfill  
  \begin{minipage}[t]{0.49\textwidth}
    \includegraphics[scale = .26]{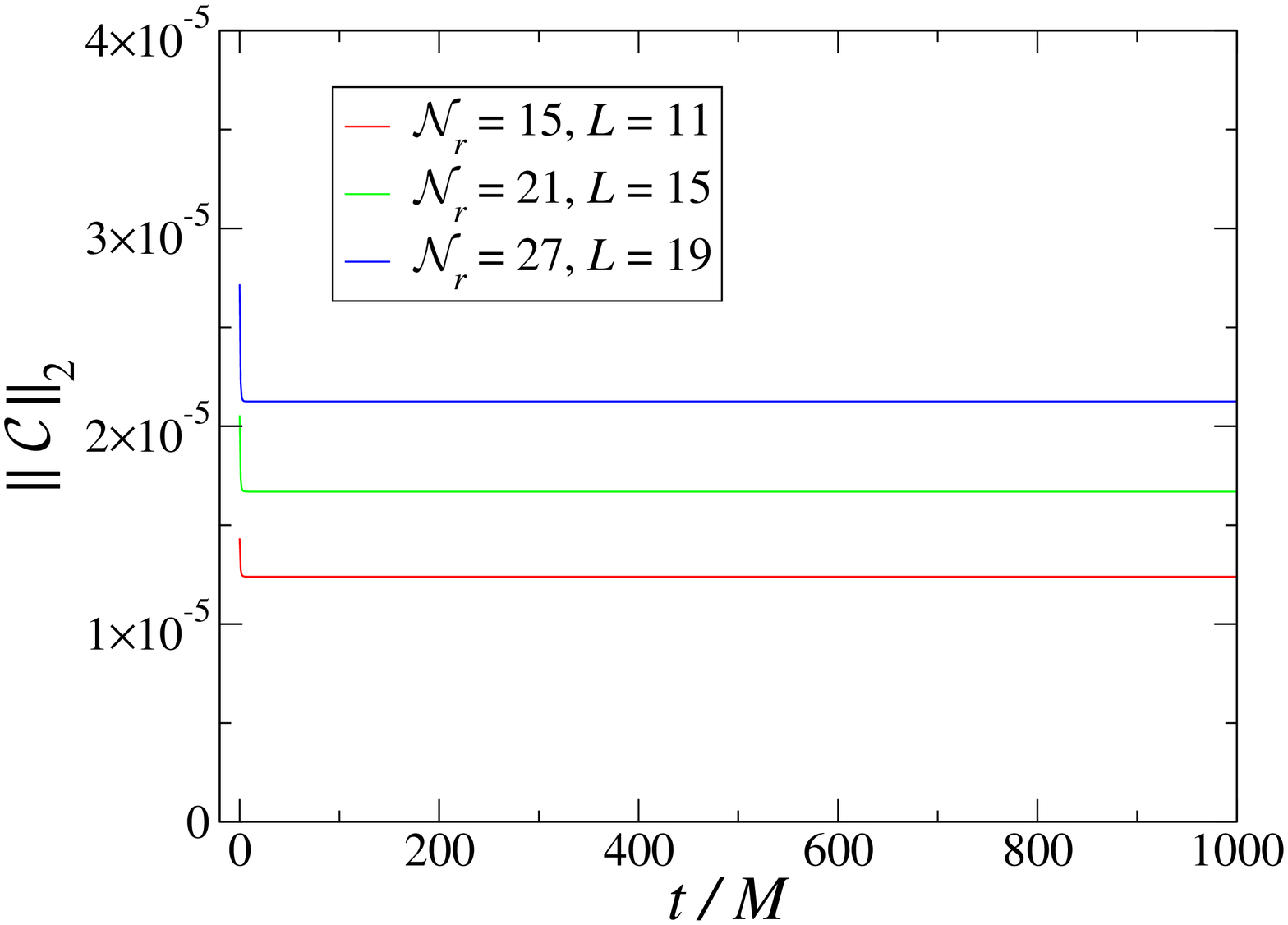}
  \end{minipage}
  \hfill  
  \begin{minipage}[t]{0.49\textwidth}
    \includegraphics[scale = .26]{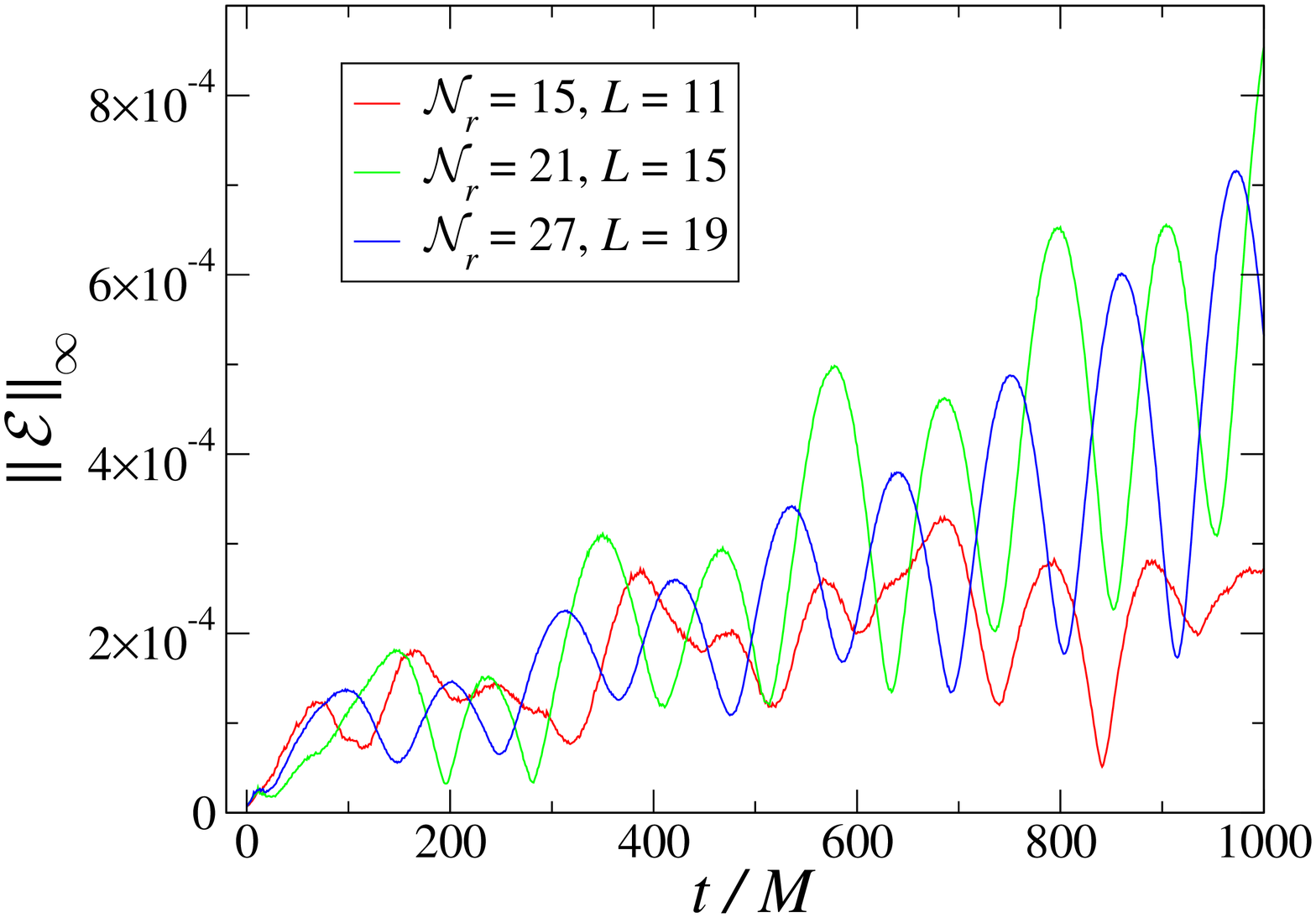}
  \end{minipage}
  \hfill  
  \begin{minipage}[t]{0.49\textwidth}
    \includegraphics[scale = .26]{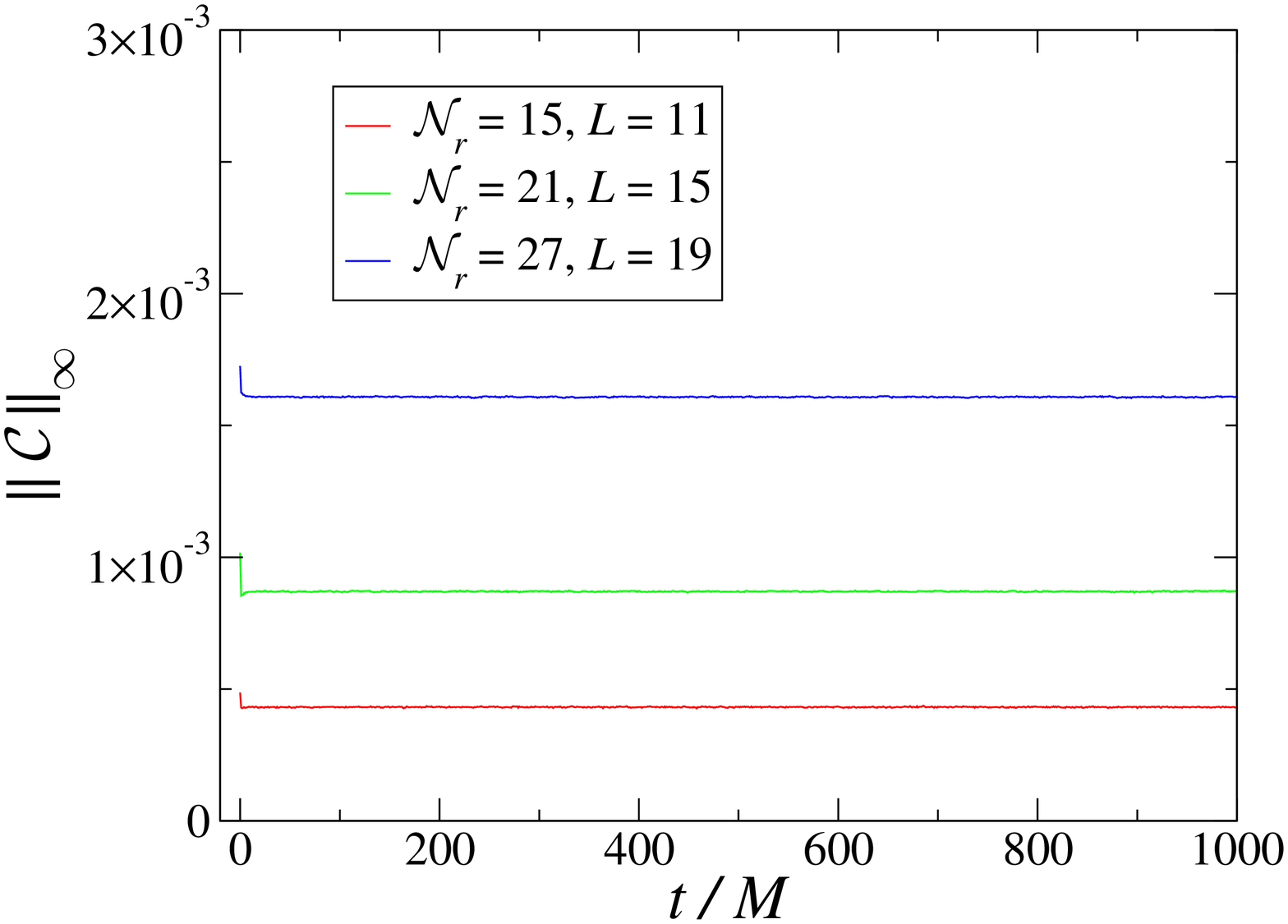}
  \end{minipage}
  \caption{\label{fig:robstabschw} 
  Robust stability test on the Schwarzschild 
  background \eref{eq:schwarzschild}. 
  $L^2$ (top) and $L^\infty$ (bottom) norms of the error (left) 
  and the constraints (right) for three different resolutions.}
\end{figure}

\begin{figure}[t]
  \begin{minipage}[t]{\textwidth}
    \centering
    \includegraphics[scale = .26]{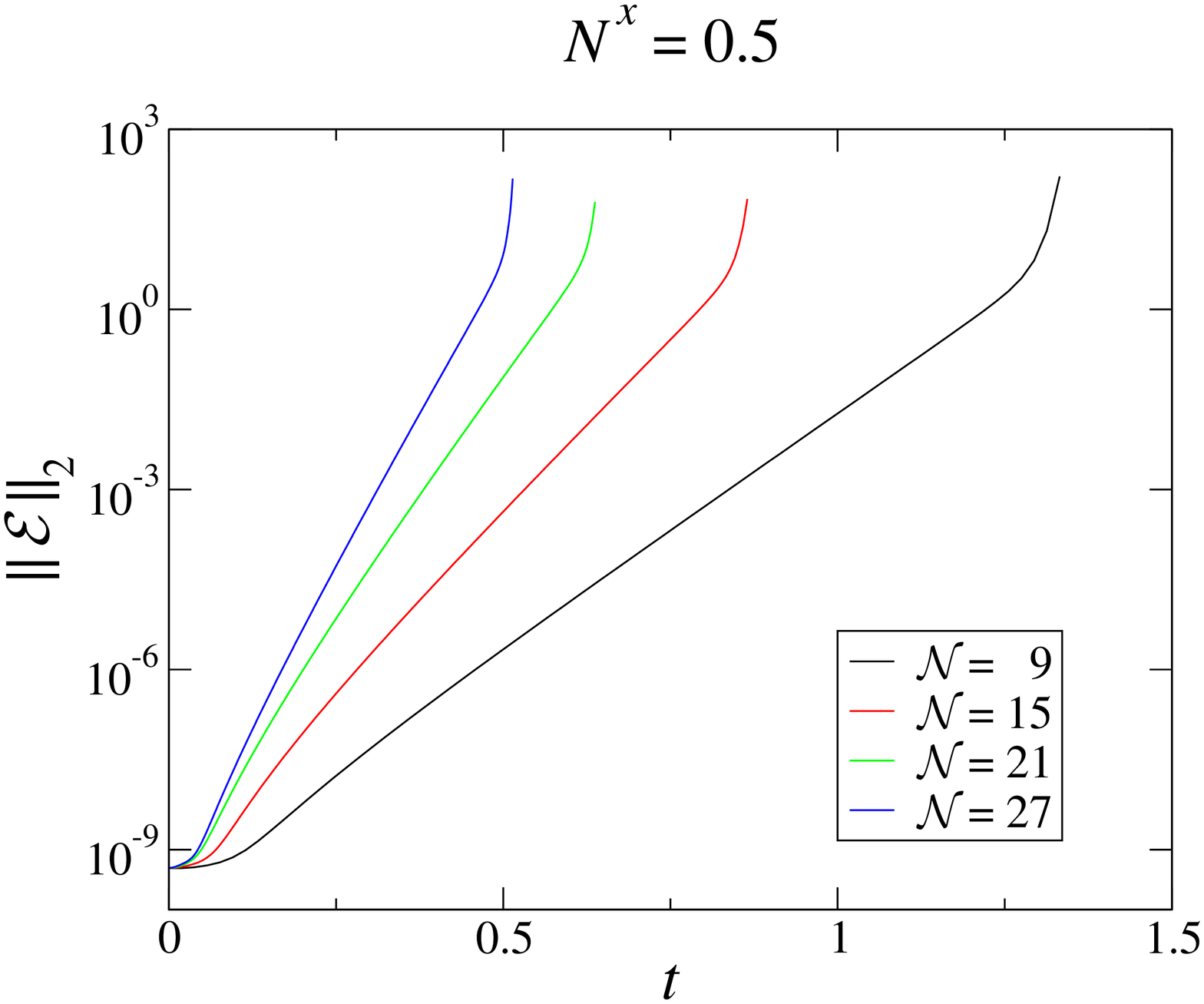}
  \end{minipage}
  \hfill  
  \begin{minipage}[t]{0.49\textwidth}
    \includegraphics[scale = .26]{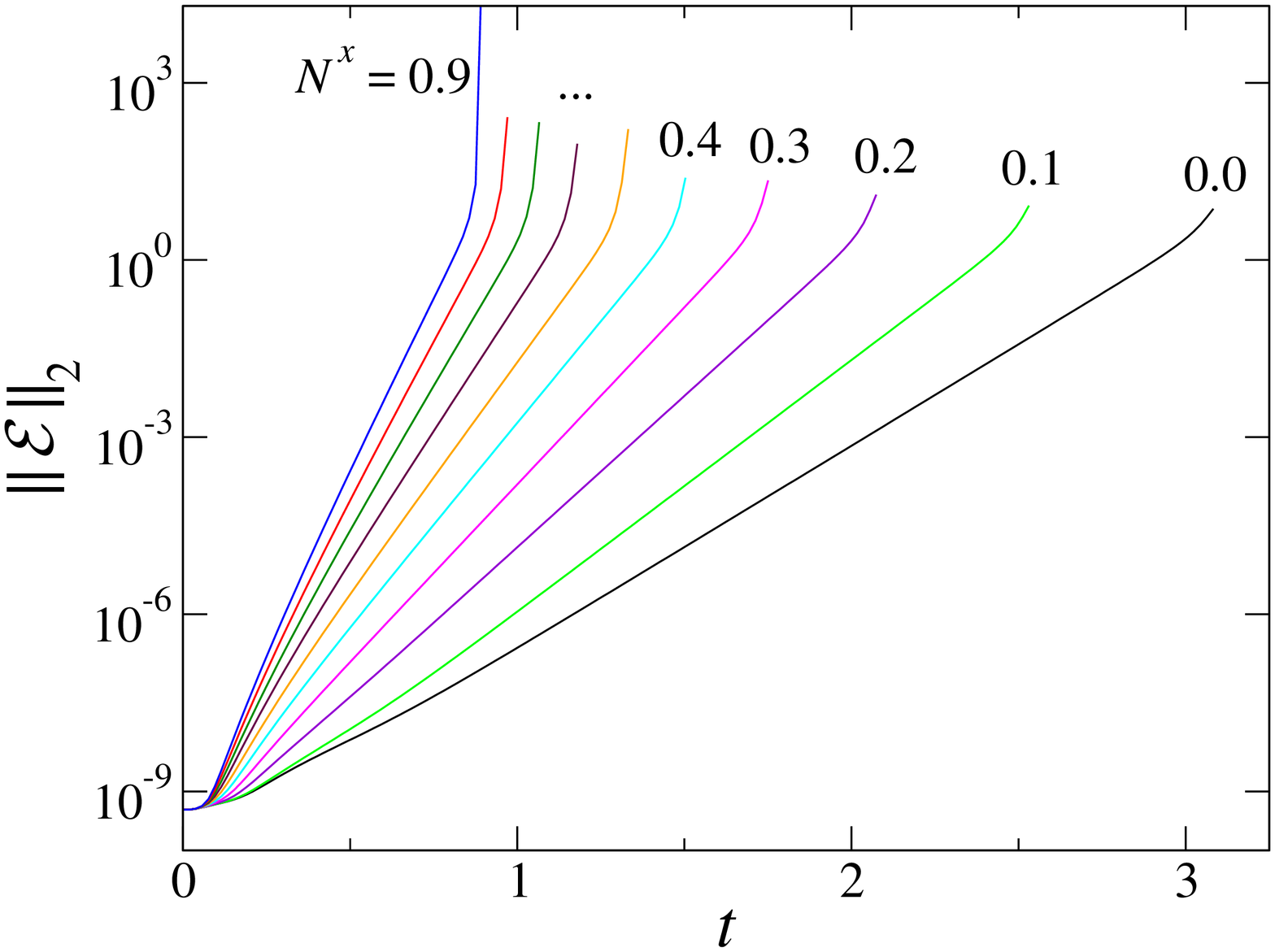}
  \end{minipage}
  \hfill  
  \begin{minipage}[t]{0.49\textwidth}
    \includegraphics[scale = .26]{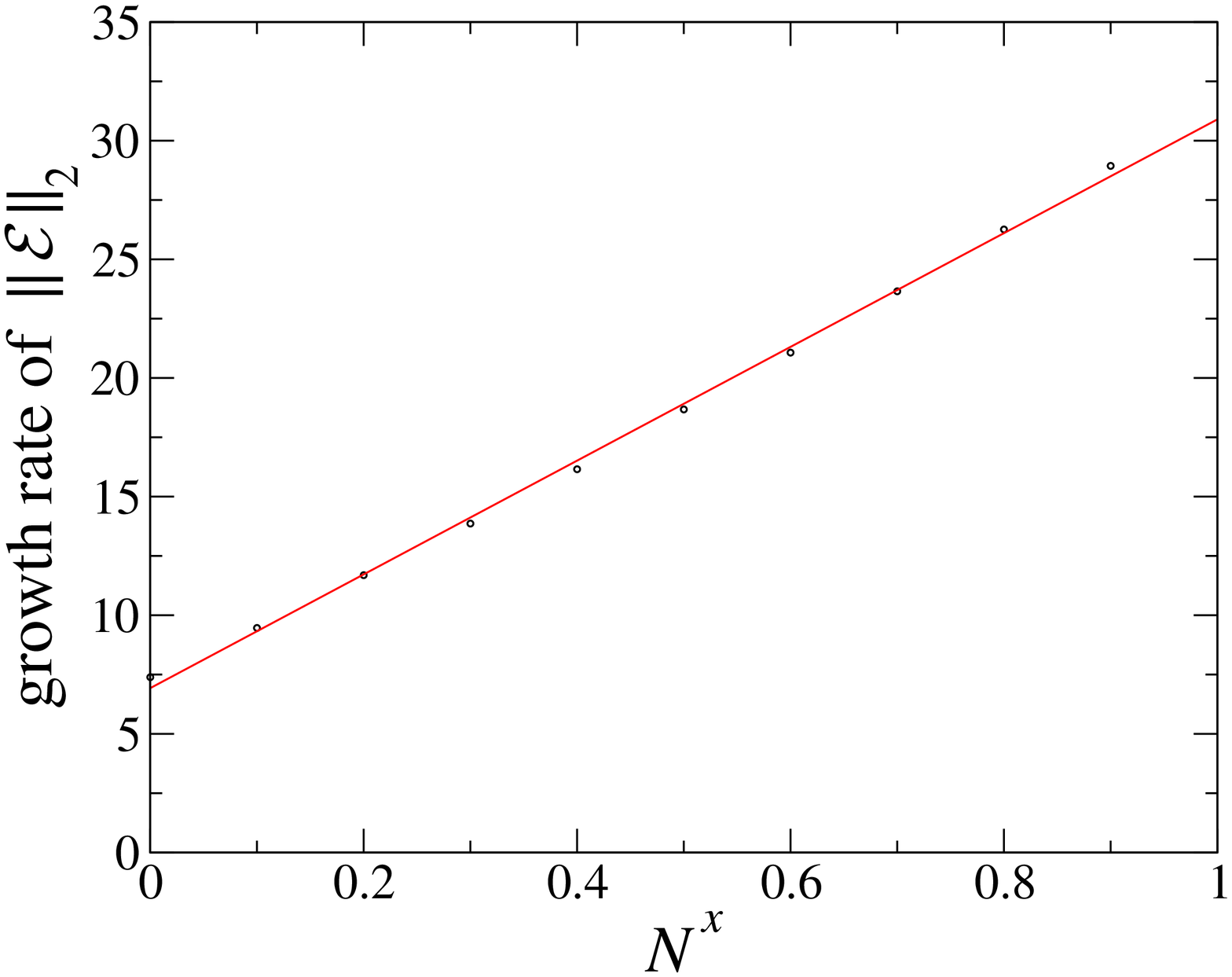}
  \end{minipage}
  \caption{\label{fig:robstabbad} 
  Robust stability test on the flat-space background \eref{eq:flatshift},
  using the alternative gauge boundary conditions \eref{eq:altgaugebc}.
  Top half: $L^2$ norm of the error for four different resolutions
  $\N \equiv \N_x = \N_y$, using $N^x = 0.5$ for the normal component
  of the shift.
  Bottom half: $L^2$ norm of the error for various values of $N^x$ at fixed
  resolution $\N  = 9$, and linear fit to the growth rate.
  (In all cases, $N^y = N^z = 0$ and $\gamma_0 = \gamma_2 = 1$.)}
\end{figure}

The following plots show the $L^2$ norms of the error
\begin{equation}
  \label{eq:error}
  \mathcal{E} = \sqrt{(\delta \psi_{ab})^2 + (\delta \Pi_{ab})^2
    + (\delta \Phi_{iab})^2}
\end{equation}
and of the constraints
\begin{equation}
  \C = \sqrt{(C_a)^2 + (\F_a)^2 + (\C_{ia})^2 + (\C_{iab})^2 + (\C_{ijab})^2}
\end{equation}
as functions of time.
In the above, the notation $(M_{ab\cdots c})^2$
refers to the sum of the squares of the Cartesian components of 
$M_{ab\cdots c}$. 
The differences $\delta$ in \eref{eq:error} are taken with respect to
the background solutions (\ref{eq:flatshift}, \ref{eq:schwarzschild}). 
When computing the quantity $\delta \psi_{ab}$, we subtract its
spatial average from it. 
This is because the constant mode of $\psi_{ab}$ typically has a large 
linear drift in time caused by a non-zero constant mode of $\Pi_{ab}$ 
(which is not eliminated by the differential boundary conditions). 
Note that this procedure does not affect the higher-frequency modes.

We begin with the flat-space background \eref{eq:flatshift}.
We run the test for two different shift vectors, $N^i = (0, 0, 0)$ 
and $N^i = (0.5, 0.5, 0)$. In the latter case, the shift points towards the
interior at the $x = -0.5$ boundary (corresponding to the $\beta > 0$
case in section \ref{sec:ghfl}) and towards the exterior at the $x = 0.5$ 
boundary (corresponding to the $\beta < 0$ case), and the shift also has a
component tangential to the boundary.
Figure \ref{fig:robstabgood} shows that in both cases, the error and the 
constraints remain of the same order as initially up to $t = 1000$ 
(and presumably forever). 
For this run we included constraint damping with parameters $\gamma_0
= \gamma_2 = 1$ (the same choice was made in \cite{GH} in order to
obtain long-term stable black hole evolutions). This leads to the sharp
initial decrease of the constraint violations (note that the randomly
perturbed initial data do not satisfy the constraints).
Because one might suspect that the constraint damping might somehow
hide potential instabilities, we rerun the test without constraint damping
(figure \ref{fig:robstabgoodnodamp}). Now the constraints grow
slightly (as expected) but still there is no sign of an instability.

Next, we turn to the Schwarzschild background \eref{eq:schwarzschild}.
Here we choose the amplitude of the random perturbations somewhat higher 
than in the flat-space case ($10^{-6}$ rather than $10^{-10}$) so that 
it is much larger than the error incurred during an evolution of 
the \emph{un}perturbed Schwarzschild spacetime for the resolutions considered.
In this case, the constraint damping is essential in order to avoid
exponential growth of the constraints \cite{GH}.
Figure \ref{fig:robstabschw} shows the results of this test, using both
the $L^2$ norm and the $L^\infty$ norm (which is more sensitive to
local effects on the boundary).
There are now significant oscillations in the error due to the much higher
amplitude of the perturbations but they grow only linearly on average
and at a rate that does not appear to increase significantly with resolution.
The constraints remain essentially constant thanks to the constraint damping.
We remark that as a consequence of the stability analysis in section
\ref{sec:ghfl}, flat space and Schwarzschild spacetime should
be equivalent with regard to stability in the high frequency limit.

The numerical results strongly suggests that the initial-boundary 
value problem is indeed well-posed, even if nontrivial initial and 
source data are included.
Previous studies \cite{SzilagyiGomezBishopWinicour,SzilagyiSchmidtWinicour,
SzilagyiWinicour,SarbachTiglio,BonaCPBC} presented similar numerical 
evidence (for different formulations), although only initial and 
boundary data but not source data were included in those robust 
stability tests.

\enlargethispage{0.5cm}
Finally, we return to flat space \eref{eq:flatshift} but now we
replace the gauge boundary conditions \eref{eq:gaugebc} 
with the alternative set \eref{eq:altgaugebc}.
The top half of figure \ref{fig:robstabbad} shows the results of the
robust stability test for shift vector $N^i = (0.5, 0, 0)$. 
Now the error grows exponentially at a rate that increases with 
resolution (note also the timescale as compared with 
figures \ref{fig:robstabgood}--\ref{fig:robstabschw}). 
This is what we expect because the initial-boundary value problem is 
ill-posed in this case (section \ref{sec:ghfl}). 
The bottom half of figure \ref{fig:robstabbad} shows what happens if
we vary the normal component $N^x$ of the shift. As predicted by the
Fourier-Laplace analysis in section \ref{sec:ghfl} (cf.~equation 
\eref{eq:badzero}), the growth rate depends linearly on $N^x$.
Numerically, we find a nonzero offset of the growth rate at $N^x = 0$,
which is not obvious from the analysis. We remark however that because
of the random perturbations, $N^x$ will never exactly vanish at the
boundary, and furthermore the argument in section \ref{sec:weakinstab}
indicates that even for $N^x = 0$, there are polynomial instabilities
of arbitrarily high polynomial order if nonzero boundary data are included.


\section{Conclusions}
\label{sec:concl}

We considered the initial-boundary value problem for a first-order
formulation of the Einstein equations in generalized harmonic
gauge. This system was derived in \cite{GH} and has proven very
successful in obtaining long-term stable black hole evolutions.
The boundary conditions we considered have the special property that
they control the incoming gravitational radiation via the incoming
fields of the Weyl tensor \cite{FriedrichNagy,BardeenBuchman,
KidderBC,SarbachTiglio}. We believe that this is essential in order
to obtain reliable information about the gravitational radiation
emitted from a compact source if the domain of integration has artificial
timelike boundaries, as is a common situation in numerical relativity.
In addition, the boundary conditions eliminate the incoming constraint fields
\cite{Stewart,IriondoReula,CalabreseLehnerTiglio,
KidderBC, BonaCPBC}, in which we believe they are superior to
constraint-preserving boundary conditions of Dirichlet 
type \cite{KreissWinicour} from a numerical point of view.

In section \ref{sec:fl}, we analyzed the well-posedness of the
initial-boundary value problem using the Fourier-Laplace technique
\cite{Stewart,CalabreseSarbach,ReulaSarbach,SarbachTiglio,RinnePhD,
  KreissWinicour}.
This required taking the high-frequency, or frozen-coefficient
approximation. To allow for an arbitrary background spacetime, we had
to take into a account an arbitrary shift vector at the boundary. This
generalizes the result stated in \cite{GH}, where only a tangential
shift was considered.
We showed that the Kreiss condition is satisfied, which implies that
the system is boundary-stable (i.e., the solution can be estimated in terms
of the boundary data).
Unlike for maximally dissipative boundary conditions, it is not known
in the present case of differential boundary conditions
whether (or under which additional assumptions) the Kreiss condition 
is also sufficient for well-posedness if non-trivial initial and source 
data are included. It would be of considerable interest to the numerical 
relativity community to obtain a general theorem covering this case.

It has been claimed \cite{ReulaSarbach,SarbachTiglio} that systems 
with differential boundary conditions
might admit weak instabilities with milder than exponential time
dependence even if the Kreiss condition is satisfied.
In section \ref{sec:weakinstab}, we considered 
instabilities with polynomial time dependence for a general first-order
initial-boundary value problem. It was found that such instabilities 
are ruled out by the Kreiss condition. (More precisely, the condition is 
that $0$ be not a generalized eigenvalue.)
For the generalized harmonic Einstein equations, it turned out that
the choice of boundary conditions for the gauge degrees of freedom can
be crucial for stability: for an innocent-looking set of alternative
gauge boundary conditions that did not satisfy the Kreiss condition,
we found a weak instability (with linear time dependence) if the
shift was tangential at the boundary. 
However, as soon as the shift pointed towards the exterior,
the weak instability was turned into a strong one with exponential time
dependence -- this demonstrates that taking into account a normal
component of the shift can be important.

Finally, we performed a numerical robust stability test \cite{AwA,Boyle,
SzilagyiGomezBishopWinicour,SzilagyiSchmidtWinicour,SzilagyiWinicour,
SarbachTiglio,BonaCPBC,Zinketal}.
The background spacetime was taken to be either Minkowski space with a
shift or Schwarzschild. We added small random perturbations to
both the initial data, the boundary conditions and the right-hand-side 
of the evolution equations.
The generalized harmonic evolution system performed very well on these tests,
with the error and the constraints remaining of the same order of magnitude
over $1000$ light crossing times in the flat-space case, or $1000 M$
in the Schwarzschild case. This strongly suggests that
the initial-boundary value problem is likely to be well-posed even 
if nontrivial initial and source data are included.
We also ran the test on the alternative set of gauge boundary
conditions, finding resolution-dependent exponential growth of the
error as predicted by the Fourier-Laplace analysis. The linear
dependence of the growth rate on the normal component of the shift was
also reproduced, although exponential growth was observed even in the
limiting case of a tangential shift.


\ack
I am particularly grateful to Olivier Sarbach for many valuable
discussions and to Mark Scheel for help with the Caltech-Cornell 
Spectral Einstein Code, which he developed jointly with Larry Kidder 
and Harald Pfeiffer.
Further I thank Lee Lindblom for suggesting this project and for
encouragement throughout the work, Alexander Alekseenko for helpful 
discussions, and all of them for careful reading of the manuscript.

This work was supported in part by a grant from the Sherman Fairchild
Foundation, by NSF grants PHY-0244906 and PHY-0601459 and NASA grants
NAG5-12834 and NNG05GG52G.

\appendix
\section*{Appendix}
\setcounter{section}{1}
\label{sec:remcon}

In this appendix, we prove a detail that we postponed in section
\ref{sec:CPBC}. Recall that in the case of an outward-pointing shift
($N^x < 0$), we could only impose constraint-preserving boundary conditions 
on the incoming constraint fields $c_a^{0-}$ and $c^4_{nAab}$.
Here we show that this implies that the remaining incoming constraint 
fields $c^2_{Aa}$ and $c^4_{ABab}$ also vanish at the boundary, 
so that we have a full set of maximally dissipative boundary
conditions for the constraint evolution system.
We restrict ourselves to the high-frequency limit and use the Fourier-Laplace
framework of section \ref{sec:ghfl}.

Suppose that we impose the constraint-preserving boundary conditions
proposed in section \ref{sec:CPBC} in the $N^x < 0$ case: in Fourier-Laplace
language, equations (\ref{eq:flc0mbc1}--\ref{eq:flgaugebc4}) with zero
data $\tilde h^{\mathrm{C}}$. These imply a linear system
\begin{equation}
  C^{(1)} \bsigma = 0
\end{equation}
for the constants $\sigma_{iab}$ parametrizing the general
Fourier-Laplace-transformed solutions \eref{eq:negbetasoln}.
The kernel of $C^{(1)}$ spans all solutions that satisfy our
constraint-preserving boundary conditions.

Next, we consider the remaining incoming constraints $c^2_{Aa}$ and 
$c^4_{ABab}$. Their Fourier-Laplace transforms are given by
\begin{eqnarray}
  \label{eq:flrcfirst}
  \tilde c^2_{y\st} &=& \partial_\xi \tilde \Phi_{yx\st}
    + \rmi(\tilde \Phi_{yy\st} + \half \tilde \Pi_{\st\st} + \half \tilde \Pi_{xx}
    + \half \tilde \Pi_{yy} + \half \tilde \Pi_{zz}) ,\\
  \tilde c^2_{yx} &=& \partial_\xi (\half \tilde \Phi_{yxx} + \half
    \tilde \Phi_{y\st\st} - \half \tilde \Phi_{yyy} - \half \tilde \Phi_{yzz}) 
    + \rmi(\tilde \Phi_{yyx} + \tilde \Pi_{\st x}) ,\\
  \tilde c^2_{yy} &=& \partial_\xi \tilde \Phi_{yxy}
    + \rmi(\half \tilde \Phi_{yyy} + \half \tilde \Phi_{y\st\st} - \half
    \tilde \Phi_{yxx} - \half \tilde \Phi_{yzz} + \tilde \Pi_{\st y}) ,\\
  \tilde c^2_{yz} &=& \partial_\xi \tilde \Phi_{yxz} 
    + \rmi(\tilde \Phi_{yyz} + \tilde \Pi_{\st z}) ,\\
  \tilde c^2_{z\st} &=& \partial_\xi \tilde \Phi_{zx\st} 
    + \rmi \tilde \Phi_{zy\st} ,\\
  \tilde c^2_{zx} &=& \partial_\xi (\half \tilde \Phi_{zxx} + \half
    \tilde \Phi_{z\st\st} - \half \tilde \Phi_{zyy} - \half \tilde \Phi_{zzz})
    + \rmi \tilde \Phi_{zyx} ,\\
  \tilde c^2_{zy} &=& \partial_\xi \tilde \Phi_{zxy}
    + \rmi (\half \tilde \Phi_{zyy} + \half \tilde \Phi_{z\st\st} - \half
    \tilde \Phi_{zxx} - \half \tilde \Phi_{zzz}) ,\\
  \tilde c^2_{zz} &=& \partial_\xi \tilde \Phi_{zxz}
    + \rmi \tilde \Phi_{zyz} ,\\
  \label{eq:flrclast}  
  \tilde c^4_{yzab} &=& \rmi \tilde \Phi_{zab} .
\end{eqnarray}
Equations (\ref{eq:flrcfirst}--\ref{eq:flrclast}) imply another linear
system for the integration constants $\sigma_{iab}$,
\begin{equation}
  C^{(2)} \bsigma = 0 .
\end{equation}

We need to check that any solution satisfying the boundary conditions 
(\ref{eq:flc0mbc1}--\ref{eq:flgaugebc4}) also satisfies equations
(\ref{eq:flrcfirst}--\ref{eq:flrclast}), in other words, that
\begin{equation}
  \ker C^{(1)} \subset \ker C^{(2)}.
\end{equation}
This is a purely algebraic condition, which is straightforward to
check with our computer algebra programme. It is indeed satisfied.


\section*{References}
\bibliographystyle{iopart-num}
\bibliography{refs}

\end{document}